\documentclass[fleqn,usenatbib]{mnras}

\usepackage{newtxtext,newtxmath}

\usepackage[T1]{fontenc}
\usepackage{ae,aecompl}

\usepackage{graphicx}	
\usepackage{amsmath}	

\usepackage{amssymb}	
\usepackage{bm}    
\usepackage{url}
\usepackage{subfig}

\usepackage{color, listings, setspace, courier, relsize}
\newsavebox{\LstBox}

\lstdefinelanguage{Python}{
    	numbers=left,
    	numberstyle=\footnotesize,
    	numbersep=7pt,
    	xleftmargi$n_{max}=$1.26em,
    	framextopmargi$n_{max}=$2em,
    	framexbottommargi$n_{max}=$2em,
    	showspaces=false,
    	showtabs=false,
    	showstringspaces=false,
    	frame=l,
    	tabsize=4,
    	stepnumber=1,
	basicstyle=\ttfamily\small,
    	backgroundcolor=\color{Background},
    	breaklines=True,
    	postbreak=\mbox{\textcolor{red}{$\hookrightarrow$}\space},
	commentstyle=\color{green}\ttfamily,
    	stringstyle=\ttfamily\color{Strings},
    	morecomment=[s][\color{Strings}]{'}{'}, 
    	stringstyle=\ttfamily\color{Comments},
    	morecomment=[s][\color{Comments}]{\#}{\#}, 			
	stringstyle=\ttfamily\color{Strings},
	morekeywords={import,from,class,def,for,while,if,is,in,elif,else,not,and,or,print,break,continue,return,access,as,except,exec,finally,global,import,lambda,pass,print,raise,try,assert},
    	keywordstyle={\color{Keywords}\bfseries}, 
    	morekeywords={[2]True,False,None},
    	keywordstyle={[2]\color{BuiltinConstant}\slshape},
	otherkeywords={[2]*},
	keywordstyle={[2]\color{Asterisk}},
	emph={self},
	emphstyle={\color{self}\slshape}	
}
   
\lstdefinelanguage{bash}{
    numbers=none,
    numberstyle=\footnotesize,
    numbersep=7pt,
    xleftmargi$n_{max}=$1.26em,
    framextopmargi$n_{max}=$2em,
    framexbottommargi$n_{max}=$2em,
    showspaces=false,
    showtabs=false,
    showstringspaces=false,
    frame=none,
    tabsize=4,
    stepnumber=1,
    basicstyle=\ttfamily\small\setstretch{1},
    backgroundcolor=\color{Background},
    breaklines=True,
    postbreak=\mbox{\textcolor{red}{$\hookrightarrow$}\space},
}

\newcommand{\ToolName}{\textsc{Lenstruction}}

\title[A versatile tool for cluster lensing source reconstruction ]{A versatile tool for cluster lensing source reconstruction. I. 
methodology and illustration on sources in the Hubble Frontier Field Cluster MACS J0717.5+3745 }

\author[L. Yang et al.]{
Lilan Yang,$^{1,2}$\thanks{E-mail: ylilan@astro.ucla.edu}
Simon Birrer,$^{3, 2}$
Tommaso Treu$^{2}$
\\
\\
$^{1}$School of Physics and Technology, Wuhan University, Wuhan 430072, China \\
$^{2}$Department of Physics and Astronomy, University of California, Los Angeles, CA 90095-1547, USA\\
$^{3}$Kavli Institute for Particle Astrophysics and Cosmology and Department of Physics, Stanford University, Stanford, CA 94305, USA
}

\date{Accepted XXX. Received YYY; in original form ZZZ}

\pubyear{2020}

\begin{document}
\label{firstpage}
\pagerange{\pageref{firstpage}--\pageref{lastpage}}
\maketitle

\begin{abstract}
 We describe a general purpose method to reconstruct the intrinsic
 properties of sources lensed by the gravitational potential of
 foreground clusters of galaxies. The tool \ToolName\ is implemented
 in the publicly available multi-purpose gravitational lensing software \textsc{Lenstronomy},
 in order to
 provide an easy and fast solution to this common astrophysical problem.
 The tool is based on forward modeling the appearance
 of the source in the image plane, taking into account the
 distortion by lensing and the instrumental point spread function
 (PSF). 
 For singly-imaged sources a global lens model in the
 format of the Hubble Frontier Fields (HFF) lensing maps is required
 as a starting point.
 For multiply-imaged sources, the tool can also
 fit and apply first (deflection), second (shear, convergence), and
 third order (flexion) corrections to the local gravitational
 potential to improve the reconstruction,
 depending on the quality of the data.  We illustrate
 the performance and features of the code with two examples of
 multiply-imaged systems taken from the Hubble Frontier Fields,
 starting from five different publicly available cluster models.
 We find that, after our correction, the relative
 magnification - and other lensing properties - between the
 multiple images become robustly constrained. Furthermore, we
 find that scatter between models of the reconstructed source size
 and magnitude is reduced. The code and jupyter notebooks are publicly available.
\end{abstract}

\begin{keywords}
galaxies:clusters:individual: MACS J0717.5+3745-- gravitational lensing: strong 
\end{keywords}



\section{Introduction}

Cluster of galaxies acts as the most powerful gravitational lenses,
magnifying and distorting background distant sources. 
For a given instrumental setup, the magnification effect enables the study of sources with higher sensitivity and resolution,
acting effectively as a cosmic telescope \citep[e.g.][]{Marshall2007}.  
Studying the background source helps to probe the galaxy formation and evolution including 
the morphology, size, kinematics, star formation history, and chemical abundances 
\citep{Richard2006, Stark2008, Sharon2012, Jones2015, Bouwens2017, Kawamata2018, deLaVieuville2019}. 
In addition, cluster lensing also contributes to understanding the mass distribution of clusters of galaxies, probing the dark matter,
the geometry and absolute scale of the universe with measurement of the time delay between multiple images 
\citep{Jullo2010,Hoekstra2013, Kelly2015, Natarajan2017, Caminha2017,Grillo2018, Birrer2019}. 

A necessary condition for exploiting scientifically the strong lensing effect is modeling the potential of the deflector.
From a technical standpoint, lens modeling of both galaxy-scale and cluster-scale lenses presents a lot of challenges.  
For example, the mass sheet degeneracy is a concern in both cases \citep{Seitz1997, Gorenstein1988}, even though the presence of multiple families of multiple images at multiple redshifts can alleviate the concern in clusters \citep{bradaclomsch2004}.
Much progress has been achieved in the past twenty
years, somehow leading to parallel and independent developments in cluster and galaxy-scale lensing.  

In galaxy-scale lensing \citep[see, e.g.,][for a review]{Treu2010},
the number of image pixels that records information of the
  lensing system is generally of order $10^4$, the deflector
is often a dynamically relaxed massive galaxy that can be described by
a relatively simple mass model, and there is one or at most two
systems of multiple images.  These features mean that one can use
computationally fast mass models \citep[e.g., the singular isothermal
ellipsoid (SIE)][]{Kormann1994} and full image plane modeling is
tractable with desktop computing power \citep{Warren2003, Treu2004,
  Vegetti2009, Tagore2014, Birrer2015, Nightingale2015}.

In contrast, in cluster-scale lensing \citep[see, e.g.,][for a
review]{Kneib2011}, the lensing potential is often complex and
multimodal, there are multiple systems of multiple images often at
different redshifts, and the image can span almost a full Hubble Space
Telescope image ($10^6$-$10^7$ pixels), albeit sparsely. 
Therefore, whereas state of the art galaxy-scale lens models can handle a full
source reconstruction, cluster lens models tend to focus on
reproducing multiple image positions and do not work at the pixel
level. A good example is provided by the Hubble Frontier Fields (HFF) lens
modeling effort. In a game-changing effort, multiple map-making teams
have made their lens models public
\footnote{\url{http://www.stsci.edu/hst/campaigns/frontier-fields/Lensing-Models}}. 
 The models are based on hitherto unprecedented numbers of multiple images, including many with spectroscopic confirmation, and have been shown to generally provide good estimates of the mass distribution and magnification effect of the clusters.  
 However, the lens model is constrained primarily by the positions of the lensed images (and sometimes weak lensing) and is not designed to do full source reconstruction on a sub-arcsecond scale \citep{Sharon2012, Meneghetti2017}.  
Furthermore, the lens-modeling teams contracted by HFF are independently using a variety algorithm resulting in differences between lens models.  
The performance of those algorithms has been investigated in the literature.  
For example, \cite{Acebron2017} and \cite{Meneghetti2017} compare the methods using simulated data, 
while \cite{Priewe2017} and \cite{Remolina2018} carried out an evaluation of lens models of the HFF cluster.
Those works have confirmed the accuracy and precision of the strong lensing methods applied, and showed that the major uncertainties in the lens models are found near cluster substructure and in the high magnification regions around the critical lines that are not immediately constrained by nearby multiple images.

Even for the most ambitious cluster-scale lens modeling projects,
additional effort \cite[e.g.,][]{Rau2014, Sharon2015} beyond image
position fitting is needed if one wants to reconstruct a specific
source in detail, and take full advantage of the quality of the data
\citep[e.g.][]{Postman2012,Treu2015, Lotz2017}. 
The difficulty of reconstruction lies in achieving lens models that are sufficiently precise for source reconstruction up to sub-arcsecond
scale, while allowing for enough freedom in the source light profiles and simultaneously dealing with blurring effect in the image plane. 
Also, the lens models often under-constrain the source-lens degeneracy for systems only observed in a single image.

In this paper, we introduce a general-purpose methodology to solve
the problem of cluster-scale source reconstruction in a variety of contexts.  
Starting with an initial guess lens model, we adopt the forward modeling approach.  
In practice, to correct the initial lens model, we employ the perturbative method proposed by \cite{Blandford2001} and further studied by \cite{Koopmans2005, Suyu2006, Suyu2009}.  
The central concept is starting from a good global lens model, then performing the localized, small-scale potential
perturbation near the particular images.  
To represent the wide range of source morphologies, we utilize a linear decomposition of the modeled source into a series of basis functions of different profiles.  
In addition, our approach uses the technology developed in the context of galaxy-scale lensing to deal with the
blurring effect from the point spread function (PSF) \citep{Treu2004, Suyu2006}.

Our approach is implemented in the code \ToolName\ powered by \textsc{Lenstronomy}\footnote{\label{lenstronomy1}\url{https://github.com/sibirrer/lenstronomy}},
a multi-purpose open-source gravitational lens modeling \texttt{python} package,
which is developed by \cite{Birrer&Amara2018}, and based on the methodology outlined by \cite{Birrer2015}.  
The scientific goal of \ToolName\ is to allow
scientists to study in detail source plane quantities like
morphologies, sizes, luminosities, star formation rates, metallicities for large sample of objects in a
self-consistent and practical way, also to explore systematic
uncertainties related to the lens models.  This need is driven by
the increasing quality and quantity of cluster-lens data from current
and future observatories, from the Hubble Frontier Fields
to e.g., the already planned guaranteed time and
  Early Release Science programs on the James Webb Space Telescope.
The software \textsc{Lenstronomy} at the core of \ToolName\, has been
applied successfully to diverse scientific problems, such as cosmographic analysis
\citep{Birrer:2016, Birrer2019, Shajib:2019H0}, modeling lensed quasars \citep{Shajib2019}, probing dark matter structure \citep{Birrer:2017substructure, Gilman2019},
quasar host galaxy decompositon \citep{Ding2019} and to generate simulations for a Convolutional Neural Network analysis (e.g. \cite{DiazRivero:2019}, Park et al. in prep, Wagner-Carena et al. in prep). 
A comparison with a different source reconstruction method by \cite{Joseph2019} is presented in their paper. We refer the reader to the GitHub repository for more general information about \textsc{Lenstronomy}.

Above applications are on the galaxy-scale regime but the same methods can be ported to the cluster regime.
As a first illustration of \ToolName, we apply it to two sets of
multiple images in the Hubble Frontier cluster MACSJ0717.5+3745,
starting from 5 different publicly available models. 
We show that relative lensing corrections are needed and substantially improve the agreement between the models.

The paper is organized as follows.  In Section \ref{method}, we review
the lens modeling technique.  In Section \ref{algorithm}, we introduce
and describe the algorithm \ToolName.  In Section \ref{sec:example},
we present two examples of source reconstruction of lensed images in
the lensing cluster MACSJ0717.5+3745.  We make comparison of relative
morphology, magnification and source properties between HFF lens models in Section
\ref{sec:comparison}.  
Summary and conclusions are given in Section \ref{sum-conclusion}.

\ToolName, together with documentation and example notebooks are publicly available  \footnote{\url{https://github.com/ylilan/lenstruction}} .
The users are kindly requested to cite this paper, \cite{Birrer&Amara2018}, and
\cite{Birrer2015}, if they make use of \ToolName.

\section{Methodology}\label{method}

\ToolName\ adopts a forward modeling approach to reconstruct source brightness distribution, simultaneously
considering lensing and blurring effects, under a Bayesian inference formalism as described Section \ref{formalism}.
We discuss the degeneracy between lensing operator and source light model, and potential correction on lensing operators in Section \ref{sec:corr-lens}.
The types of descriptions of the surface brightness distribution of the source currently implemented in \ToolName\ are given in Section \ref{sec:corr-source}.
The technique adopted to regularize model complexity is described in Section \ref{sec:bic}.

\subsection{Forward modeling technique: source to image plane mapping}\label{formalism}

In a forward modeling approach, to reconstruct the source surface brightness, $\bm{S}$, from the data, $\bm{D}$, we first use a theoretical light profile of $\bm{S}$ and predict the lensed image(s) $\bm{D}$ as  
\begin{equation}\label{eq-modeldata}
\bm{D} = \bm{B} \cdot \bm{L} \cdot \bm{S},
\end{equation}
where $\bm{L}$ is the lensing operator, and $\bm{B}$ is the PSF blurring operator that is determined by the property of the telescope. 
Following \cite{Birrer2015}, for a given lensing operator, the scaling of surface brightness in the source results in a linear response of the model. We thus impose a set of basis function describing the source suface brightness, $\bm{S}_i$, with a vector of flux normalization coefficients, $\xi$, such that the total source is a linear superposition of those basis set. 
The modeled image is then a linear superposition too with the response matrix $X = \bm{B} \cdot \bm{L} \bm{S}_i$ such that the modeled image $\bm{D_m} = X\xi$.

We employ Bayesian inference methods to estimate the posterior distribution function, $p(\bm{D}_m|\bm{D}_o)$, of the free parameters in the model.
The log-likelihood function of the observed data $\bm{D}_o$ given a model $\bm{D}_m$ is,

\begin{equation}\label{likelihood}
\begin{split}
\log p(\bm{D}_o|\bm{D}_m) &=\sum_{i=1}^{N_d}\frac{(D_{o,i}-D_{m,i})^2}{2\sigma_{i}^2}+\text{const}
\end{split}
\end{equation}
where $D_{o,i}$ and $D_{m,i}$ are the observed and modeled flux in each pixel respectively, $\sigma_i$ is the error in each pixel and
$N_d$ is the total number of pixels in the modeled image. We estimate the error in each pixel as a combination of a Gaussian background rms, $\sigma^2_{\rm bkg}$ and a Poisson term scaled with the exposure time, $f_i$, of each pixel, see details in \cite{Birrer2015} as 
\begin{equation} 
\sigma^2_i = (\sigma^2_{\rm bkg} + D_{m,i}/f_i).
\end{equation}
The best-fitting parameters of source light model is estimated by maximizing the posterior distribution function. The source configuration for a given lensing operator and basis set can be solved by a linear minimization problem:
\begin{equation} \label{wls}
	\bm{\xi_0} = \text{min} \|W^{1/2}(\bm{D}_o-\bm{D}_m)\|^2 =  \text{min}_{\bm{\xi}} \|W^{1/2}(\bm{D}_o-X\xi)\|^2 .
\end{equation}
where W is the weight matrix. In our case, treating noise as uncorrelated, $W$ is the diagonal matrix with $W_i = \sigma_i^{-2}$.
In practice, we utilize Weighted Least Square (WLS) method to reconstruct the source surface brightness.

To figure out the best-fit configuration of the source
light model, we adopt two steps.  First, we need to find a solution
for a given lens and source model assumptions. Then, we change the
choices of model complexity as long as changing and increasing
complexity improves the results.  

To simplify the problem, the effects of dust and the contamination by
foreground lens light have been ignored in this paper. Taking dust
and foreground light into account in the forward modeling is
straightforward from a conceptual point of view and implemented in \textsc{lenstronomy} \citep[see e.g.][]{Shajib:2019H0}.

\subsection{Degeneracies and observational constraints on the lensing operator}\label{sec:corr-lens}

\subsubsection{Degeneracies}\label{sec:degeneracy}

The unknown intrinsic source inherits a degeneracy with the lensing operator $\bm{L}$.
As shown in Figure~\ref{fig:multi}, let us assume that we observe image $\bm{D}_{o1}$, and the initial guess of the
lensing operator is $\bm{L}_1$. Given the data and the lensing operator, we can reconstruct the source $\bm{S}$. 
We can introduce any arbitrary correction operator $\bm{J}$ provided that there exists a corresponding inverse $\bm{J^{-1}}$ with $\bm{J}\cdot\bm{J^{-1}}$ equals the unit operator to transform simultaneously the lensing operator $\widetilde{\bm{L}} \equiv \bm{L}\cdot \bm{J}$ and the source $\widetilde{\bm{S}}\equiv \bm{J^{-1}} \cdot \bm{S}$
 without any observable effect on the data as
\begin{equation}\label{eq-single}
\begin{split}
\widetilde{\bm{D}}_{m1}& =\bm{B}\cdot \widetilde{\bm{L}}_1 \cdot \widetilde{\bm{S}}\\
&=\bm{B}\cdot (\bm{L}_1 \cdot \bm{J}) \cdot (\bm{J^{-1}} \cdot \bm{S}) \\
& =\bm{B}\cdot \bm{L}_1 \cdot \bm{S}=\bm{D}_{m1} .
\end{split}
\end{equation}
Unless intrinsic knowledge of the source is assumed or available through other means,
it is impossible to tell if the true source is $\bm{S}$ or $\widetilde{\bm{S}}$. The equation above effectively describes the most general lensing degeneracy of which the mass sheet transform \citep{Falco:1985} is the special case where $\bm{J}$ is a scalar.

\subsubsection{Constraints in multiply-imaged case} \label{sec:constraint}

\begin{figure} 
    \includegraphics[width=\columnwidth]{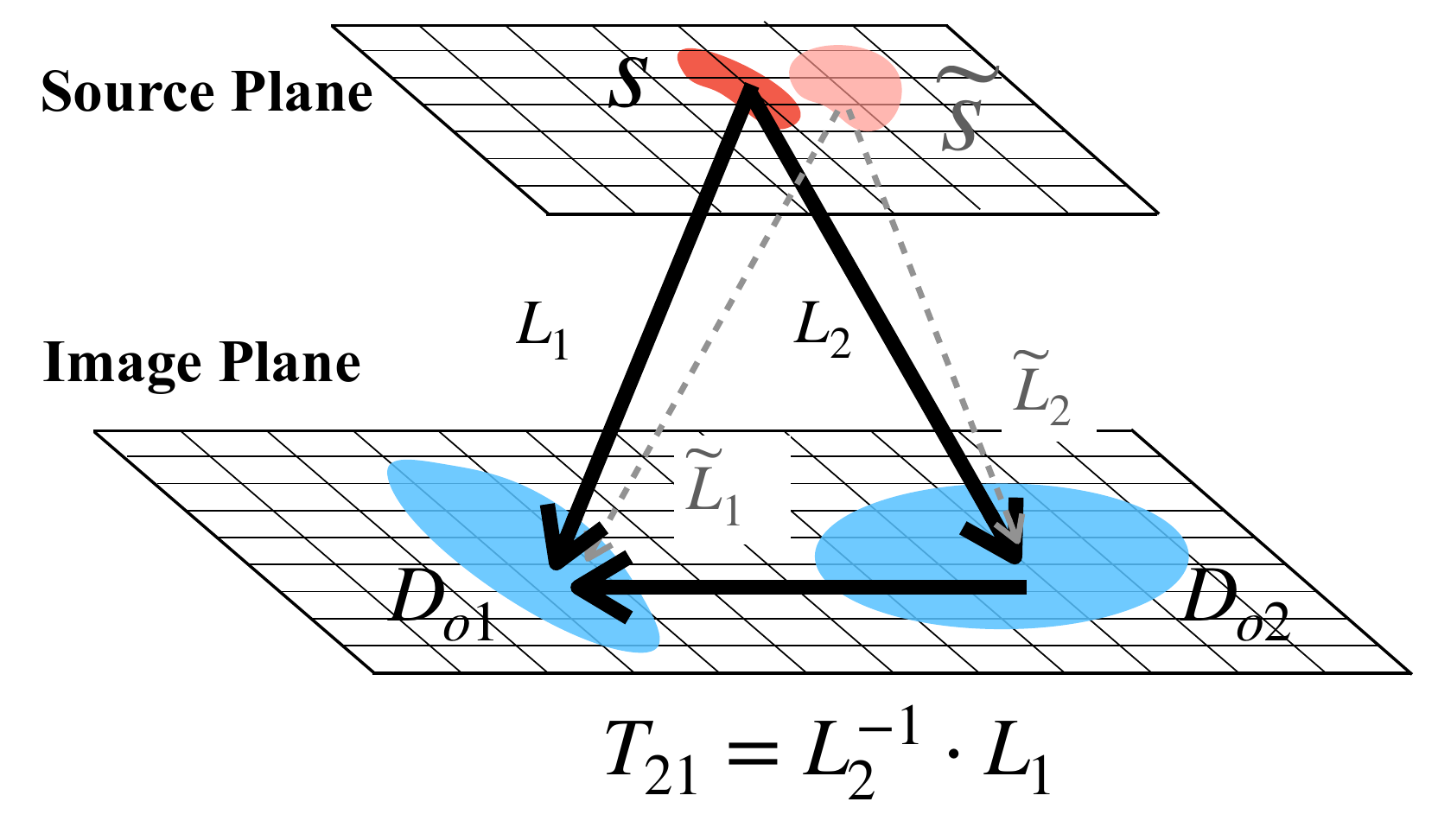} 
    \caption{Source to image plane mapping, where the source $\bm{S}$ is lensed by a foreground object, and 
     $\bm{L}_1$ and $\bm{L}_2$ are lensing operators for lensed images $\bm{D}_{o1}$ and  $\bm{D}_{o2}$ respectively. 
    The figure illustrates the degeneracies and constraints in a strong lensing system. 
    The degeneracies arise because if $\bm{L}_1$ is corrected to $\widetilde{\bm{L}}_1$, the observables are unchanged if the source is corrected to $\widetilde{\bm{S}}$ (likewise if $\bm{L}_2$ is corrected to $\widetilde{\bm{L}}_2$.) 
    The constraints arise in the multiply-imaged case.  The operators $\bm{L}_1$ and $\bm{L}_2$ are related through the transformation matrix $\bm{T}_{21}$.
 }
    \label{fig:multi}
\end{figure}

When multiple images are available of the same source, one can obtain information of the \textit{relative} lensing operators, even though of course the absolute intrinsic source properties are mathematically unknown due to the degeneracy described in Section \ref{sec:degeneracy}.
In the case of two images as described in Figure~\ref{fig:multi}, the two modeled lensed images of $\bm{D_{o1}}$ and $\bm{D_{o2}}$ are
\begin{equation}\label{d2d1}
\begin{split}
\bm{D}_{m1} =  \bm{B}\cdot \bm{L}_1  \cdot \bm{S} \\
\bm{D}_{m2} =  \bm{B}\cdot \bm{L}_2  \cdot \bm{S}  .
\end{split}
\end{equation}
Two images from the same source $\bm{S}$ are related by the transformation operator $\bm{T}_{21}$ mapping image 2 to image 1.  
\begin{equation}\label{t21}
\bm{D}_{m2} \to\bm{D}_{m1}:  \bm{T}_{21}= \bm{L^{-1}}_2 \cdot \bm{L}_1,
\end{equation}
where $\bm{T}_{21}$ is independent on a source distortion operator $\bm{J}$ and the constraints on $\bm{T}_{21}$ 
solely depends on the quality and information of the observations \citep[see e.g.,][]{Wagner2019}. 
$\bm{T}_{21}$ is the primary operator that needs to be sufficiently accurate to allow a simultaneous reconstruction of both images.
If the initial lensing operator $\bm{L^{-1}}_2 \cdot \bm{L}_1$ is insufficiently accurate, the WLS fails to reconstruct a source that matches both images simultaneously:
\begin{equation} \label{wls-multi}
\bm{\xi_0} = \text{min} \| W_i^{1/2}(\bm{D}_{oi}-\bm{D}_{mi})\|^2_{1,2} =  \text{min}_{\bm{\xi}} \| W_i^{1/2}(\bm{D}_{oi}-X_i\xi)\|^2_{1,2}
\end{equation}
where i = 1, 2 for two images respectively and the norm simultaneously applied for the same coefficients $\xi$ on both images.
Thus, in the case of multiple images, we can constrain the lensing operators based on information contained in the transformation matrix $\bm{T}_{21}$.
For example, the magnification ratio predicted by lensing operators is intuitively well constrained by the observed images.
In practice, the amount of corrections to be applied to the initial lensing operators depend on the quality of the data.
In this work we restrict ourselves to third order polynomial perturbations of the lensing potential, which are sufficient for the vast majority of sources behind cosmic telescopes that do not suffer from extreme distortion. Correction to higher-order are needed for highly distorted sources.
If the highly distorted feature appears in comparison of multiple lensed images, higher order lensing effect can be constrained using a more general approach departing from a Taylor series expansion in the lensing potential (see details in Birrer 2020, in preparation).

\subsection{Model surface brightness distribution of the source}\label{sec:corr-source}

In our approach, the source surface brightness distribution is a linear combination of a set of simple models.
To describe a wide range of unknown background astronomical sources with a finite sets, we make use of elliptical S\'ersic \citep{Sersic1968} and the two-dimensional Cartesian shapelets \citep{Refregier2003} \footnote{\textsc{Lenstronomy} supports a variety of non-linear profiles, and \ToolName\ allows for the full support of the available functionality of \textsc{Lenstronomy} }.
 Shapelets are given by:
\begin{equation}
	\bf{B}_{\bf{n}}(x;\beta_{s}) \equiv \beta_{s}^{-1}\phi_{n_1}(\beta_{s}^{-1}x_1) \phi_{n_2}(\beta_{s}^{-1}x_2),
\end{equation}
where $\beta_{s}$ is a characteristic scale, $\phi_{n_1}$ and $\phi_{n_2}$ are one-dimensional Cartesian shapelet, as:
\begin{equation}
	\phi_n(x) \equiv \left[2^n\pi^{\frac{1}{2}}n! \right]^{-\frac{1}{2}} H_n(x) e^{-\frac{x^2}{2}}
\end{equation}
where $n$ is the order of $H_n$, the Hermite polynomial.
The order $n_{\rm max}$ determines numbers of basis sets m by $m=(n_{\rm max}+1)(n_{\rm max}+2)/2$.
As the order increases, one can capture more complexity in the source surface brightness profile.
The characteristic scale $\beta_{s}$ is typically about the size of the source.
The minimum and maximum scales being resolved up to order $n$ are given by $sl_{\rm min}=\beta_{s}/\sqrt{(n_{\rm max}+1}$ and $sl_{\rm max}=\beta_{s}\sqrt{(n_{\rm max}+1)}$.

\subsection{Model complexity regularization}\label{sec:bic}

For a given model complexity, the source is reconstructed via linear minimization. 
We then repeat the procedure while varying model complexity, 
and we employ the Bayesian information criterion (BIC) method 
to balance goodness of fit and model complexity simultaneously following \cite{Birrer2019}. BIC is computed as 
\begin{equation}
\label{eq:bic}
{\rm BIC}={\rm ln}(N_d)N_k-2{\rm ln}(\hat{L})
\end{equation}
where $N_d$  and $N_k$ are the number of data points and free parameters within the model respectively, 
and $\hat{L}$ is the maximum likelihood value given the model.
Usually, the likelihood increases with the source light model complexity.
However, the number of free parameters also increases, and the minimum BIC criterion balances the increase in source complexity with the additional parameters in order to avoid over-fitting the data.

\section{Overview of  \ToolName }\label{algorithm}

To facilitate the forward modeling approach described in Section~\ref{method} and its applicability in the cluster regime on real data to the broader community, we develop the \texttt{python} package \ToolName\  build on top of \textsc{Lenstronomy}. 
\textsc{Lenstronomy} provides the core functionalities of the modeling and fitting described in Section \ref{method} and is the work horse underneath through which those tasks are executed.
\ToolName\ is the layer on top that provides the interface to the specific cluster data products and executes the specific tasks required to achieve reliable source reconstructions and lens model corrections in the cluster lensing regime.

\ToolName\ contains several independent modules.
The core module of \ToolName\ is \texttt{ClsrWorkflow} (Cluster lensing source reconstruction Workflow). \texttt{ClsrWorkflow} inherits the \texttt{Workflow} module of \textsc{Lenstronomy} \citep{Birrer&Amara2018}
tailored for perturbative lens modeling and source reconstruction in the cluster context and manages fitting and sampling routines as described in Section~\ref{sec:clsrworkflow}. The linear minimization, exploration of parameters space, and model complexity regularization are all performed through this module.
The other modules are described in detail in Section~\ref{sec:fit-prep}.
The \texttt{DataProcess} module configures the imaging data to be modeled by the \texttt{ClsrWorkflow}.
The \texttt{LensSpecify} module handles the configuration of the lens model and the \texttt{SourceSpecify} module handles the configuration of the source model to be passed into the core \textsc{Lenstronomy} modules.
\ToolName\ inherits conventions and many functionalities from \textsc{Lenstronomy} and allows to keep up with the development of the broader  \textsc{Lenstronomy} ecosystem.

\subsection{Configuration of the data and model setup}\label{sec:fit-prep}
The \texttt{DataProcess} module manages and facilitates the retrieval of the relevant information of the lensed images from the data, such as the blurring operator, known in astronomy as the point spread function (PSF), positional information and the coordinate system, pixel size, exposure time and noise, and casts those quantities into the conventions used by \textsc{Lenstronomy}.
The identification of lensed image makes use of $detect\_sources$ and $deblend\_sources$ in package $photutils$ \citep{Bradley2019} \footnote{\url{https://photutils.readthedocs.io/en/stable/index.html}}.
For the PSF, both pixelized convolution kernels, as well as analytic profiles, are supported through \textsc{Lenstronomy}. 
This is a clear step forward with respect to most previous work on extended images in cluster lensing, in which the blurring effect is often ignored.

The \texttt{LensSpecify} module defines the parameterization of the lensing operators and sets up the fitting configuration.
In the current implementation, 
we assume that the lens potential is approximately smooth over the area spanned by each observed image.
Thus, the initial shear and convergence $(\gamma_1, \gamma_2, \kappa)$ are taken directly from the convergence and shear maps provided in the input setting.
Higher-order flexion terms $( \mathcal{F}_1,\mathcal{F}_2,\mathcal{G}_1, \mathcal{G}_2)$ can also be initialized when required. The default flexion terms are set to zero,
assuming that the initial model is insufficiently accurate to provide valid initial guesses.

Users can specify the parameters to be held fixed, and assign bounds and priors to the free parameters during the modeling procedure.

The \texttt{SourceSpecify} module provides the functionality to describe the surface brightness in the source plane with various analytic profiles as well as representations in shapelet basis sets in conjunction with the \textsc{Lenstronomy} \texttt{LightModel} module.
The superposition of profiles is allowed, and the user can choose whether the constraints of the independent profile types are connected or not.

\subsection{Modeling management}\label{sec:clsrworkflow} 

The \texttt{ClsrWorkflow} module is designed to model a
wide range of source and lens model complexities in the cluster environment.

\begin{figure*}
	\includegraphics[width=1.2\columnwidth]{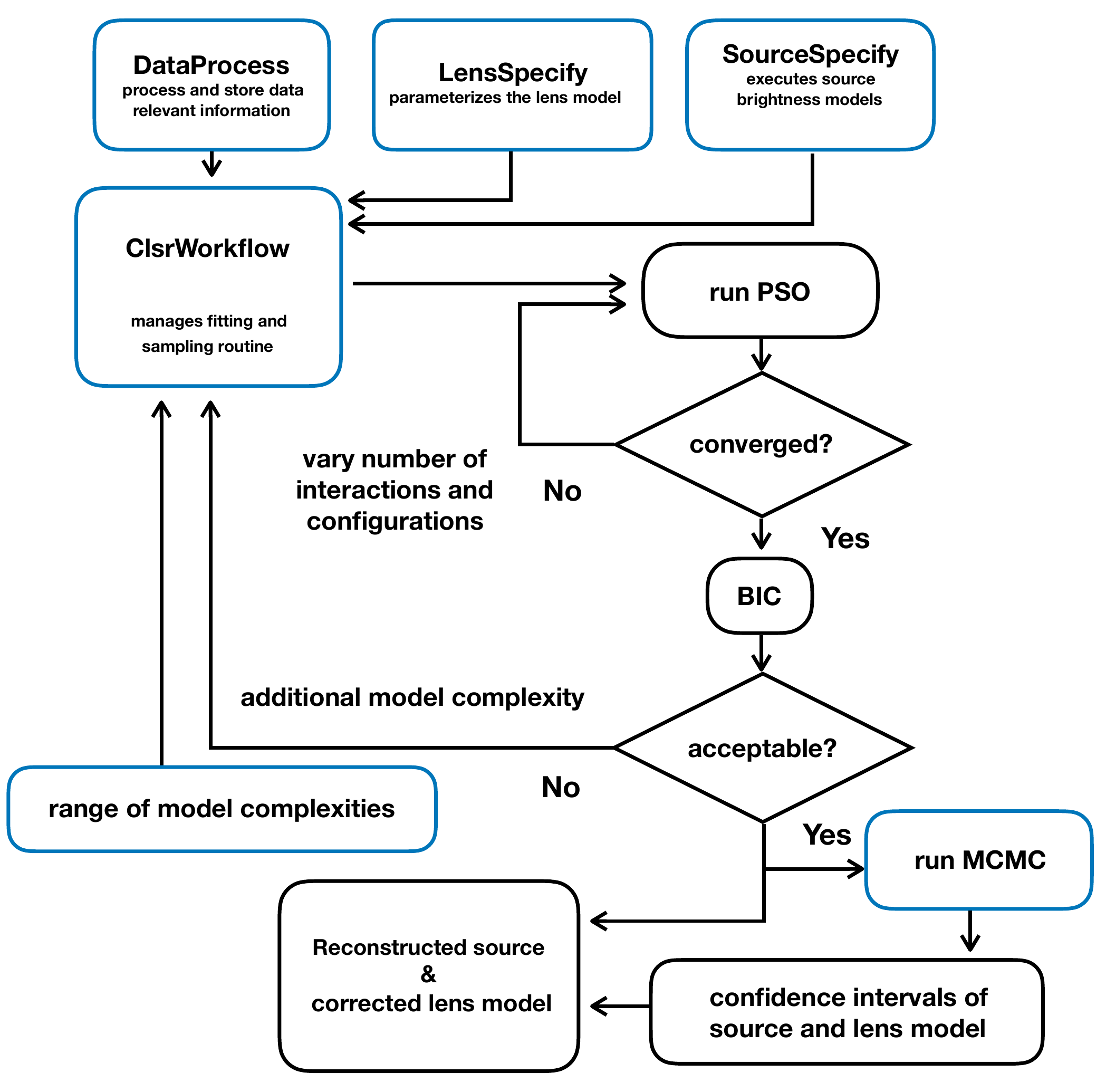}
    \caption{Flowchart of \ToolName. It consists of four modules \texttt{DataProcess}, \texttt{LensSpecify},  \texttt{SourceSpecify}, and  \texttt{ClsrWorkflow}. 
    \texttt{DataProcess} manages the pre-processing of data and storage of relevant information, e.g., data selection, exposure time, pixel size, etc. 
    Then \texttt{LensSpecify} parameterizes the lens model from the available lens model, and \texttt{SourceSpecify} specifies light profile description of the background source.
    Next, \texttt{ClsrWorkflow} takes care of fitting and sampling. Within the range of model complexities specified by the user, PSO explores the parameters space
    to figure out the solution while BIC regularizes the complexity. In the end, \ToolName\ returns the reconstructed source and corrected lens model
    for multiply-imaged cases. MCMC is also available to explore the confidence intervals of the source and lens model, once the optimal degree of complexity is selected.
    User interaction is needed for the steps in blue, while automated tasks are shown in black. }
    \label{fig:workflow}

\end{figure*}

The \texttt{ClsrWorkflow} module operates, as shown in Figure~\ref{fig:workflow}. 
It starts with the lowest model complexity, and default setting is elliptical $S\'ersic$ for source light model and
lensing parameters up to convergence and shear. For a given lensing
operator and source brightness distribution, it then solves for the
source parameters via linear minimization, 
see details in Equation~\ref{wls}.
Changes in lensing
operator and/or source brightness distribution, however, require
solving a nonlinear problem.  This step is
performed by a Particle Swarm Optimization (PSO) \citep{Kennedy1995}.  PSO
optimizes a candidate solution by employing particles to the explore
parameter volume. The particles are expected to swarm toward the best solution. 
The use of multiple particles is aimed at avoiding local maxima as it is often the case of optimizers starting from a single point.

If the adopted models can not produce an acceptable fit of the data,
the module increases the model complexity. Higher-order lensing corrections
or shapelets are included and the fitting is repeated. The process is
repeated until the minimum BIC is reached. Once the minimum BIC
solution is found, one can run a more time-consuming Markov chain
Monte Carlo (MCMC) process to explore the full posterior and provide
confidence intervals as well as degeneracies among model parameters
\citep[using emcee][]{Foreman-Mackey2013}.

\section{Examples: analysis of two multiply-imaged systems in the lensing cluster MACSJ0717.5+3745}\label{sec:example}

As an illustration of \ToolName\ and to demonstrate the power of the modules and their underlying algorithm on a real and complex example, 
we present source plane reconstruction of two multiply-imaged sources in one of the HFF cluster MACSJ0717.5+3745 \citep{Ebeling2014} at redshift $z_{\rm lens}=0.545$.
The pre-processing of the images and set-up of model configurations are illustrated in Section \ref{sec:data-model-pre}.
\ToolName\ is applicable for both singly- and multiply- imaged sources. 
We present the details of reconstruction Section \ref{sec:bradac}.

\subsection{Data and model configuration} \label{sec:data-model-pre}

\begin{figure*}
\centering
	\includegraphics[width=1.3\columnwidth]{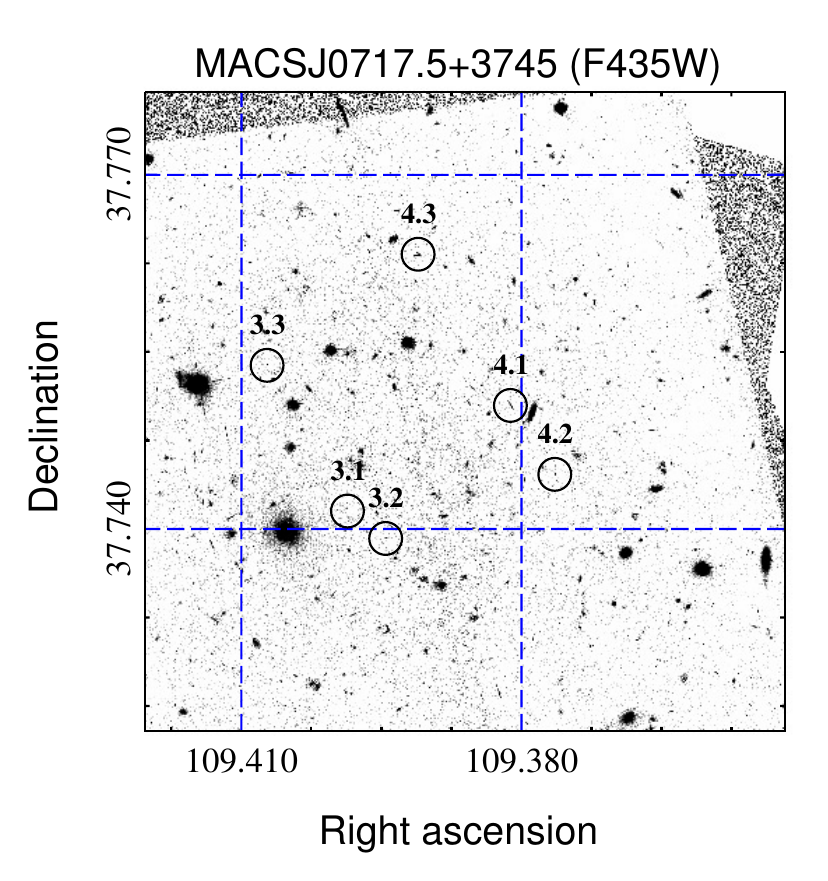}
	    \caption{HST (F435W) image of MACSJ0717.5+3745. The circles show the positions of the lensed images (details in Table~\ref{table:imgsets}.) 
	    Note how the lensed images are contaminated by foreground emission and span a large area, so that one needs to make efficient use of the pixels carrying information to optimize computational resources.  
	    }
    \label{fig:cluster_img}
\end{figure*}

\begin{figure*}
\centering
	\includegraphics[trim = 60mm 90mm 30mm 70mm, clip,  width=2\columnwidth]{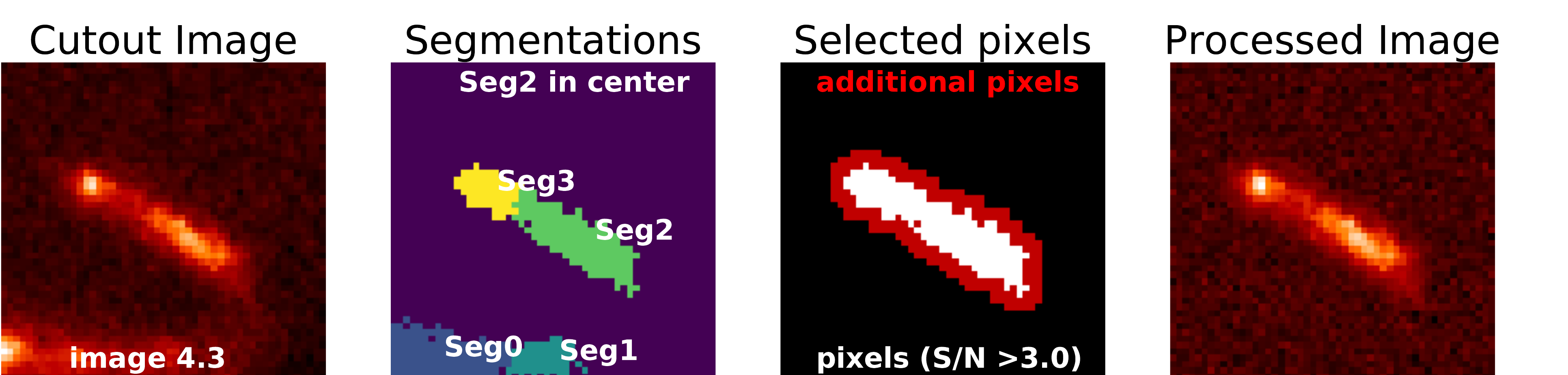}
   	
\caption{\ToolName\ steps for identifying the lensed images. 
The first column shows the cutout of lensed image 4.3. 
Pixels with S/N > 3 are identified and de-blended into segmentations maps as shown in the second column.
In the third column, we select "Seg2" and "Seg3", together with pixels surround that covers pixels with relative lower S/N, are identified as lensed image.
The fourth column shows processed image.
Note: \ToolName\ also enables the user to take contamination emission, e.g., "Seg 0" and "Seg1", into consideration, e.g., fitting it as a S\'ersic.} 
    \label{img:cleaned-map4}
\end{figure*}

Hubble Space Telescope images (through filter F435W) of two multiply-imaged sets are shown in Figure~\ref{fig:cluster_img}. 
Coordinates are listed in Table~\ref{table:imgsets}.
For consistency with previous work, we keep the same IDs of the two systems (3 and 4) as in the paper by \citet{Schmidt2014}.
The \texttt{DataProcess} module can deblend multiple images from potential foreground contamination, as shown in Figure~\ref{img:cleaned-map4}.
For this example, a bright star with coordinate  (RA, Dec) = (109.3778185, 37.75322111) is selected as the fiducial PSF.

\begin{table}
	\centering
	\caption{Multiply-imaged galaxies in the lensing cluster MACSJ0717.5+3745. The first column lists the naming convention for multiply-imaged system. The other columns list right ascension, declination (J2000) and redshift 
(more details are given by \citep{Schmidt2014}).}	
	\begin{tabular}{lccc} 
		\hline
		\hline
		 ID & RA & DEC & redshift\\
		\hline
		3.1&   109.398545&  37.741498   &  1.855  \\
		3.2&  109.394459 &  37.739172   &  1.855  \\
 	         3.3&  109.407156 &  37.753831   & 1.855    \\
		4.1&   109.381093 &  37.750440   &  1.855  \\
 		4.2&   109.376338 &  37.744602   & 1.855   \\
		4.3&   109.391097 &  37.763077   & 1.855 \\
 	 		\hline
	\label{table:imgsets}	
	\end{tabular}
\end{table}

We adopt lens models from five independent teams contracted by HFF, Brada\v{c}, Williams, CATS,  Zitrin, and Sharon \citep{ Bradac2005, Bradac2009, Williams2016Sebesta, Liesenborgs2007, Limousin2016, Zitrin2015,
Sharon2014Johnson}, to initialize \texttt{LensSpecify} modules.
Table~\ref{table:models} provides a summary of the models.
Values of shear, convergence $(\gamma_1, \gamma_2, \kappa)$ are reported in Table~\ref{table:initial-sys3} and Table~\ref{table:initial-sys4} for two lensed systems respectively.

\begin{table}
	\centering
	\caption{Lensing models utilized of the cluster MACSJ0717.5+3745. }	
	\begin{tabular}{ccc} 
		\hline
		\hline
		Model & Version &  Method \\
		\hline
		Brada$\v c$ & v1 & Pixellated\\
		Williams & v4.1 & Pixellated\\
		 CATS & v4.1 & Simply-parametrized\\   
		 Zitrin-ltm & v1 & Simply-parametrized\\
		 Sharon & v4 & Simply-parametrized\\
		 \hline
	\label{table:models}	
	\end{tabular}
\end{table}

\subsection{Description of modeling procedure} 
\label{sec:bradac}

We now describe in detail the \texttt{ClsrWorkflow} module for the two real multiply-imaged case studies. 
We first show results starting from the Brada\v{c} model, 
then the comparison to the outcome using a different model as initialization is given in Section \ref{sec:morph-comp}.

\subsubsection{Modeling details of the multiply-imaged system 3} \label{sec:sys3}

\begin{figure*}  
 \textbf{Singly-imaged cases with uncorrected lens parameters }\par\medskip
{\includegraphics[width=1.4\columnwidth]{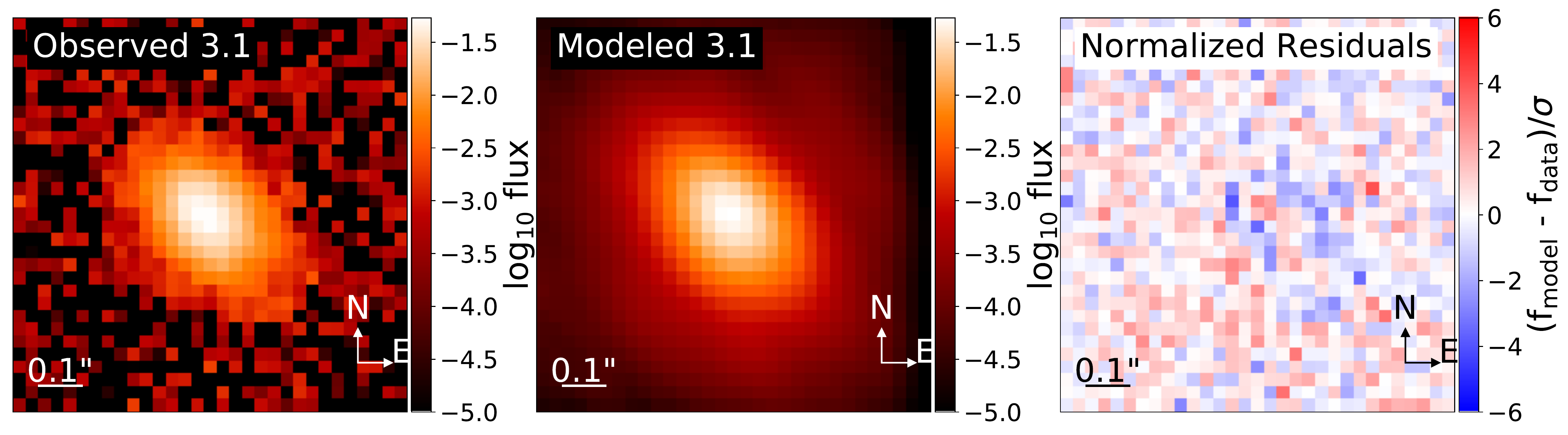} }
{\includegraphics[trim = 20mm 16mm 20mm 10mm, clip, width=0.6\columnwidth]{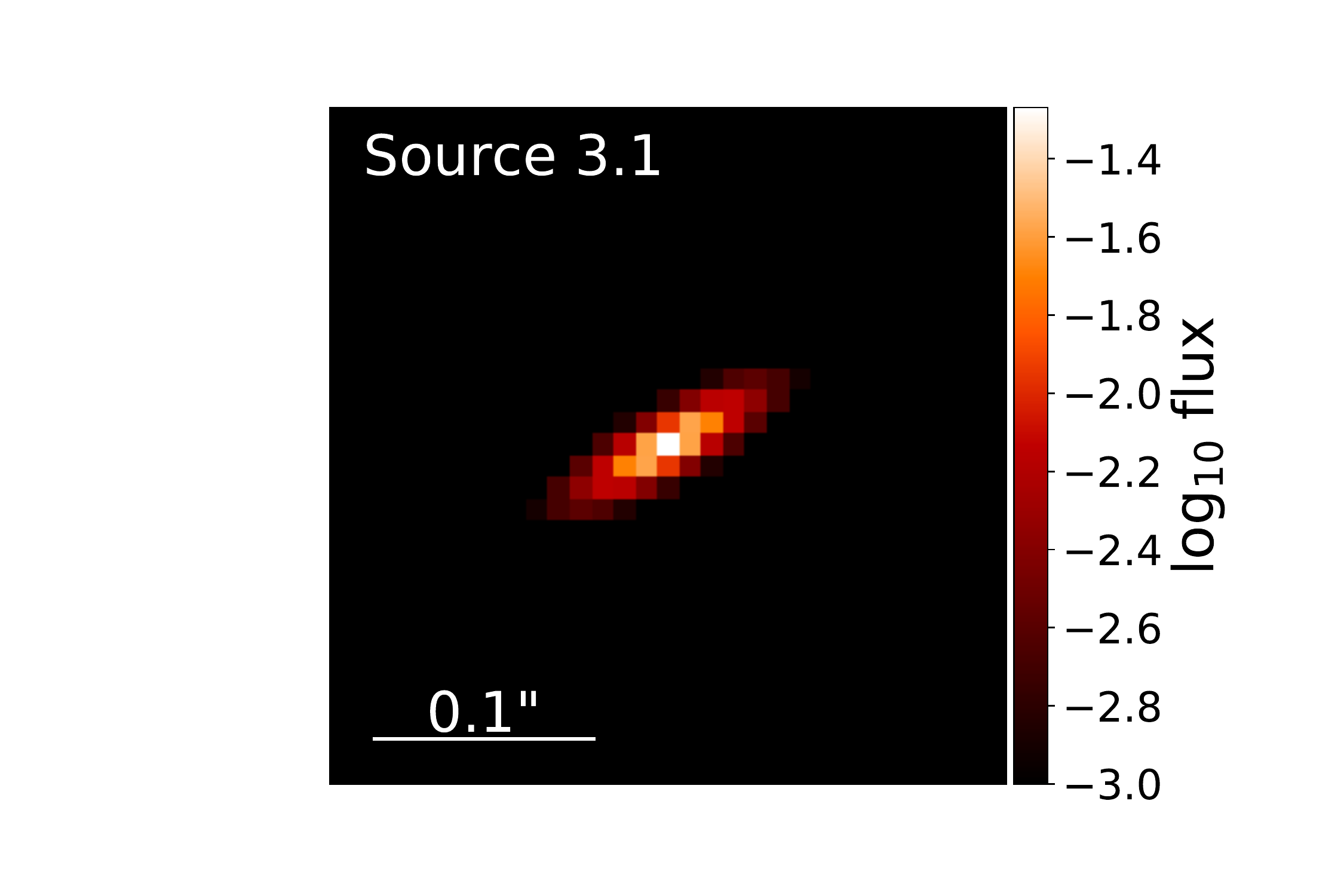} }\\
{\includegraphics[width=1.4\columnwidth]{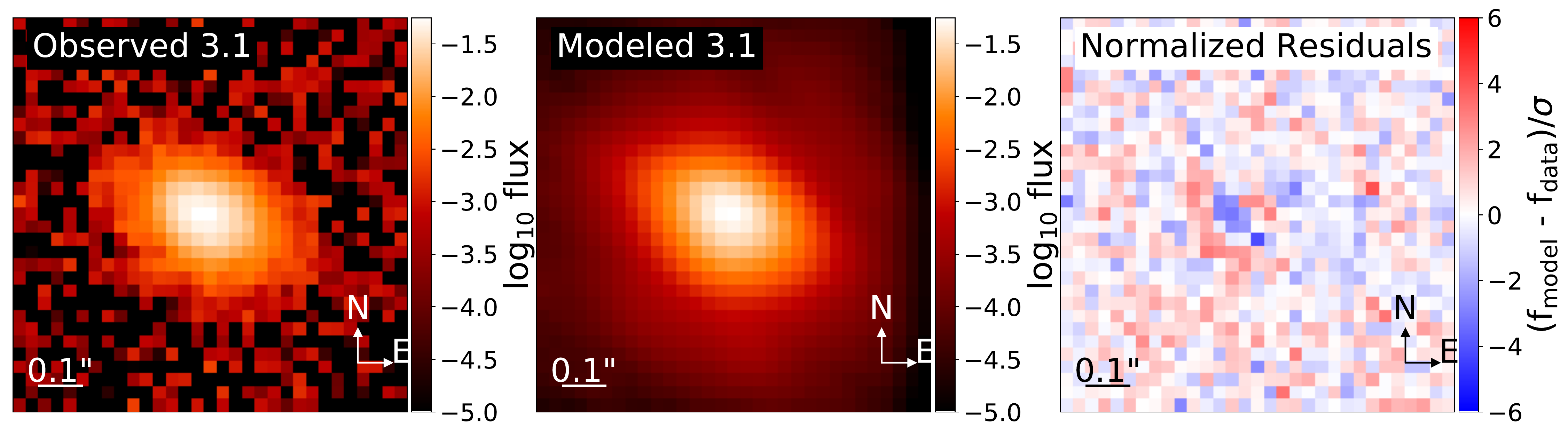} }
{\includegraphics[trim = 20mm 16mm 20mm 10mm, clip, width=0.6\columnwidth]{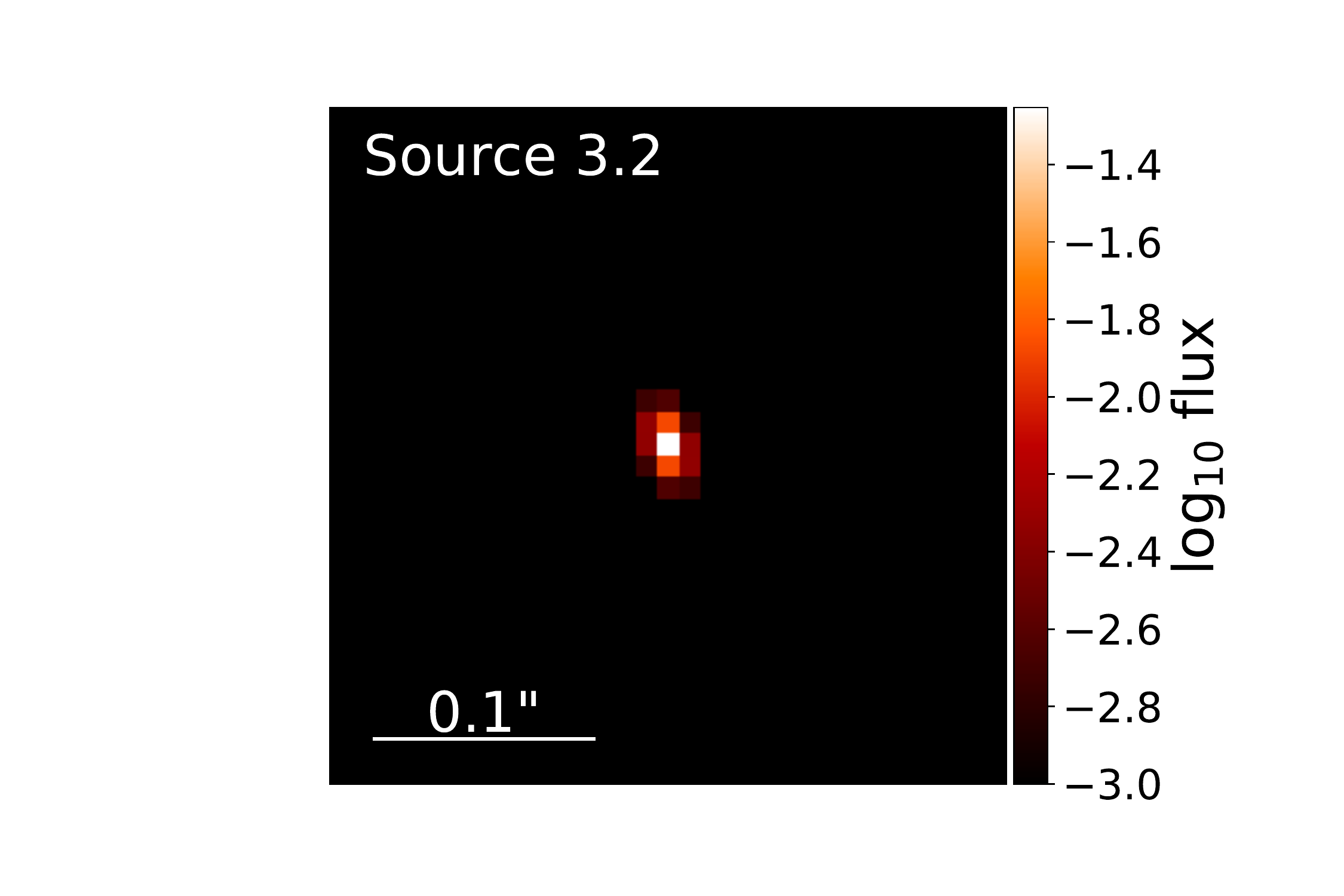}}\\

\subfloat[From left to right, we show the observed lensed images, the modeled lensed images, the normalized residuals (i.e., divided by uncertainty) and the reconstructed sources.
The reconstructed sources appear significantly different and even those are expected to be the same, illustrating the expected limitations of global models in reproducing the local potential.  Note: the reconstructed sources are re -centered on their flux centroids, to correct the errors in deflection angles of the global models shown in  Figure~\ref{fig:b}.]{\includegraphics[width=1.4\columnwidth]{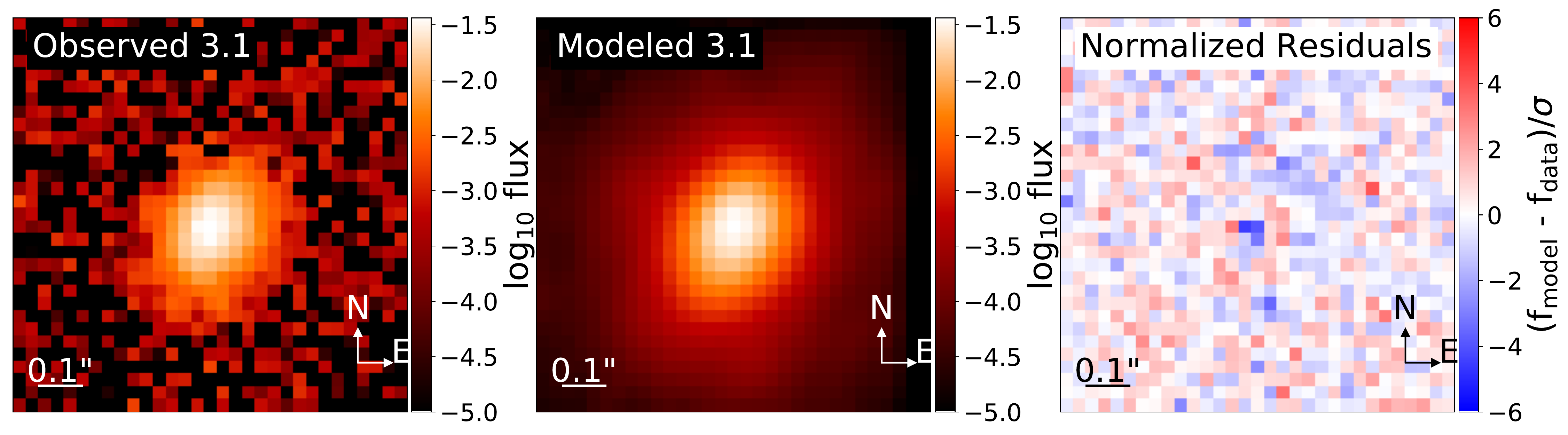} 
\includegraphics[trim = 20mm 16mm 20mm 10mm, clip, width=0.6\columnwidth]{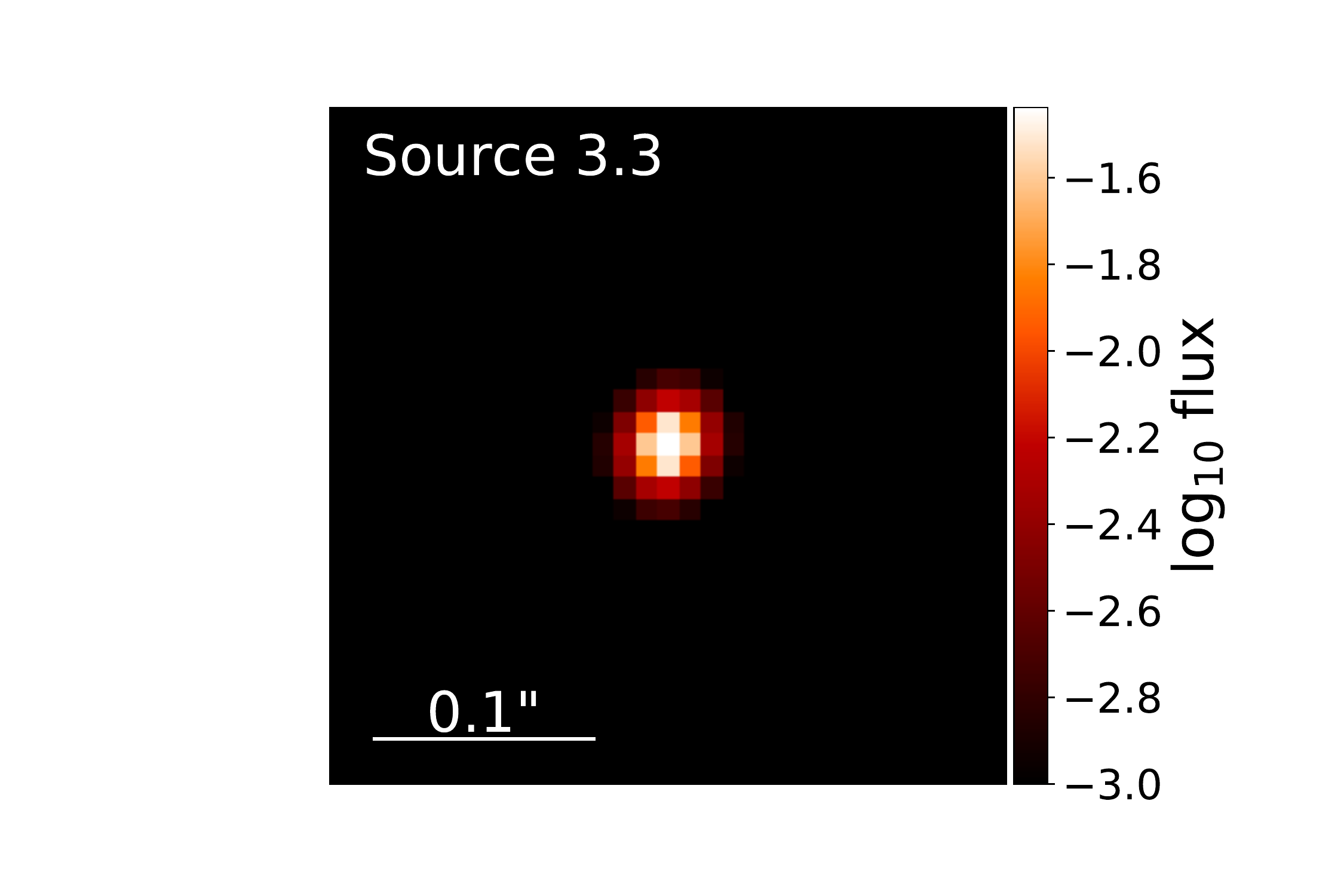} \label{fig:a} }\\

 \subfloat[Example of positional offsets of  the multipled images traced back to the source plane based on the initial model. These offsets in deflection angles are corrected in our approach by recentering.]{\includegraphics[width=\columnwidth]{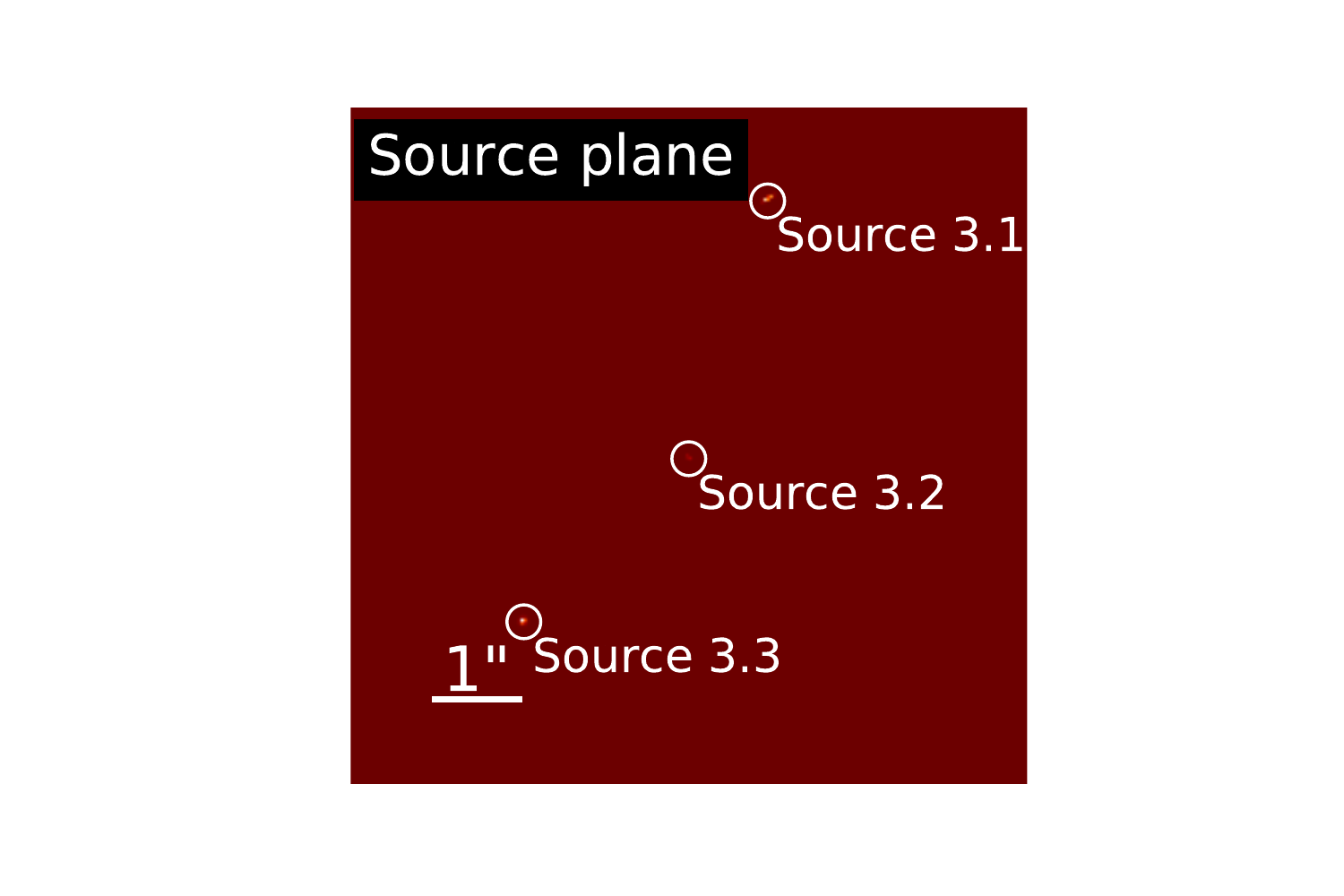} \label{fig:b} }
 
\caption{Demonstration of the modeling results of three singly-imaged cases with uncorrected lens model from the Brada$\v c$ team.}
   \label{fig:single}
\end{figure*}

To start with a simpler problem, we consider each lensed image within multiply-imaged system 3 as an individual singly-imaged source,
i.e., doing the source reconstruction independently, not demanding a joint source morphology.
The morphologies of the observed images are compact, 
thus we just apply the lowest model complexity, an elliptical S\'ersic profile for source, convergence and shear for the lens model.
The singly-imaged system lacks the information required to constrain the lensing operator see details in Section \ref{sec:degeneracy}, so we fix the lens parameters after initialization.
We present the sources reconstructed from each image in Figure~\ref{fig:single}, while fitting results of this exercise are in Table~\ref{table:initial-sys3}.

\begin{figure*}
 \centering
  \textbf{Multiply-imaged system 3 with uncorrected lens parameters }\par\medskip
 \subfloat[Same as the first three columns in Figure \ref{fig:a} (first order (deflection) has been corrected), but the residuals are much more significant. ]{\includegraphics[width=1.8\columnwidth]{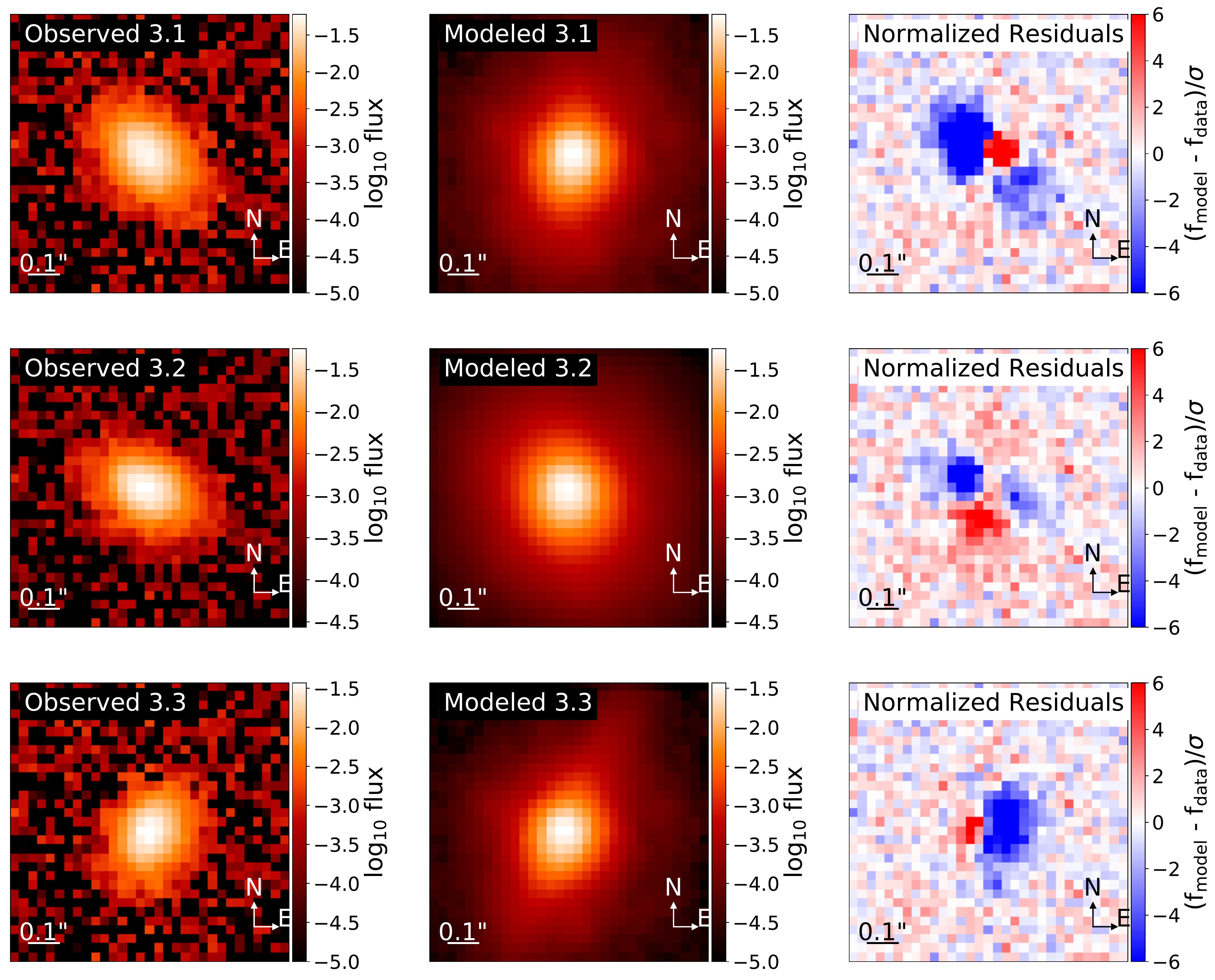} }\\
 \subfloat[Reconstructed source surface brightness distribution via uncorrected lens models from Brada$\v c$.]{ \includegraphics[width=\columnwidth]{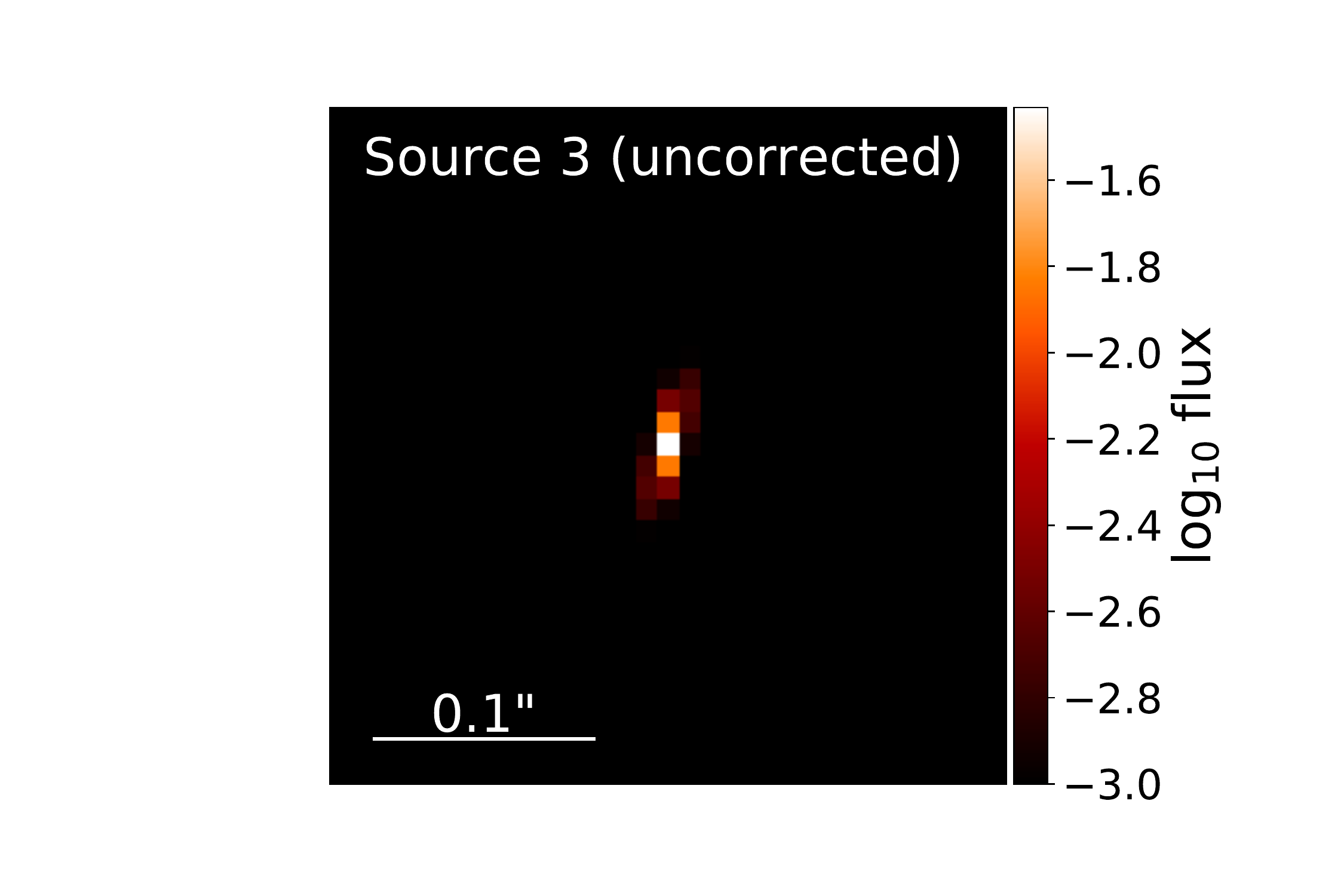} }
  
\caption{Demonstration of the modeling results of the multiply-imaged system 3 with uncorrected lens model from the Brada$\v c$ team.}
      \label{fig:uncorrected-recon3}
\end{figure*}

For  a multiply-imaged system, observed multiple images provide information to constrain the relative lensing operator (see details in Section \ref{sec:constraint}).
Before correcting the lens model, we present the results obtained using the uncorrected lens model for the combined reconstruction of the multiple images.
As shown in Figure~\ref{fig:uncorrected-recon3}, the uncorrected lens model leads to poor results. 
The initial lens model is not sufficiently accurate to match all images simultaneously at the pixel level and needs corrections as expected.
As only relative lens parameters can be constrained, letting all lensing operators free will unavoidably lead to degeneracies. 
To avoid this pitfall, we fix the lens parameters of the least magnified image (image 3.3 is the least magnified evaluated from  Brada\v{c} team, see Table~\ref{table:initial-sys3})  to the value estimated by the global model. It is important to use the least magnified image because the uncertainties of magnification are proportional to magnification itself \citep{Meneghetti2017}. Also, the current implementation of our code only applies corrections up to flexion, and therefore it is best to take the least distorted image as reference. The full underlying degeneracy inherent in lensing can then be reconstructed analytically from the reconstruction and posteriors.

\begin{figure*}
 \centering
   \textbf{Multiply-imaged system 3 with corrected lens parameters }\par\medskip

 \subfloat[Same as the first three columns in Figure \ref{fig:a} for the Brada\v{c} team.
The residuals illustrate the improvement in the fit compared with the uncorrected initial lens model, shown in Figure~\ref{fig:uncorrected-recon3} .   
However, there are still significant residuals, especially for Image 3.3, indicating that the lens model is not yet sufficiently complex.
]
{\includegraphics[width=1.8\columnwidth]{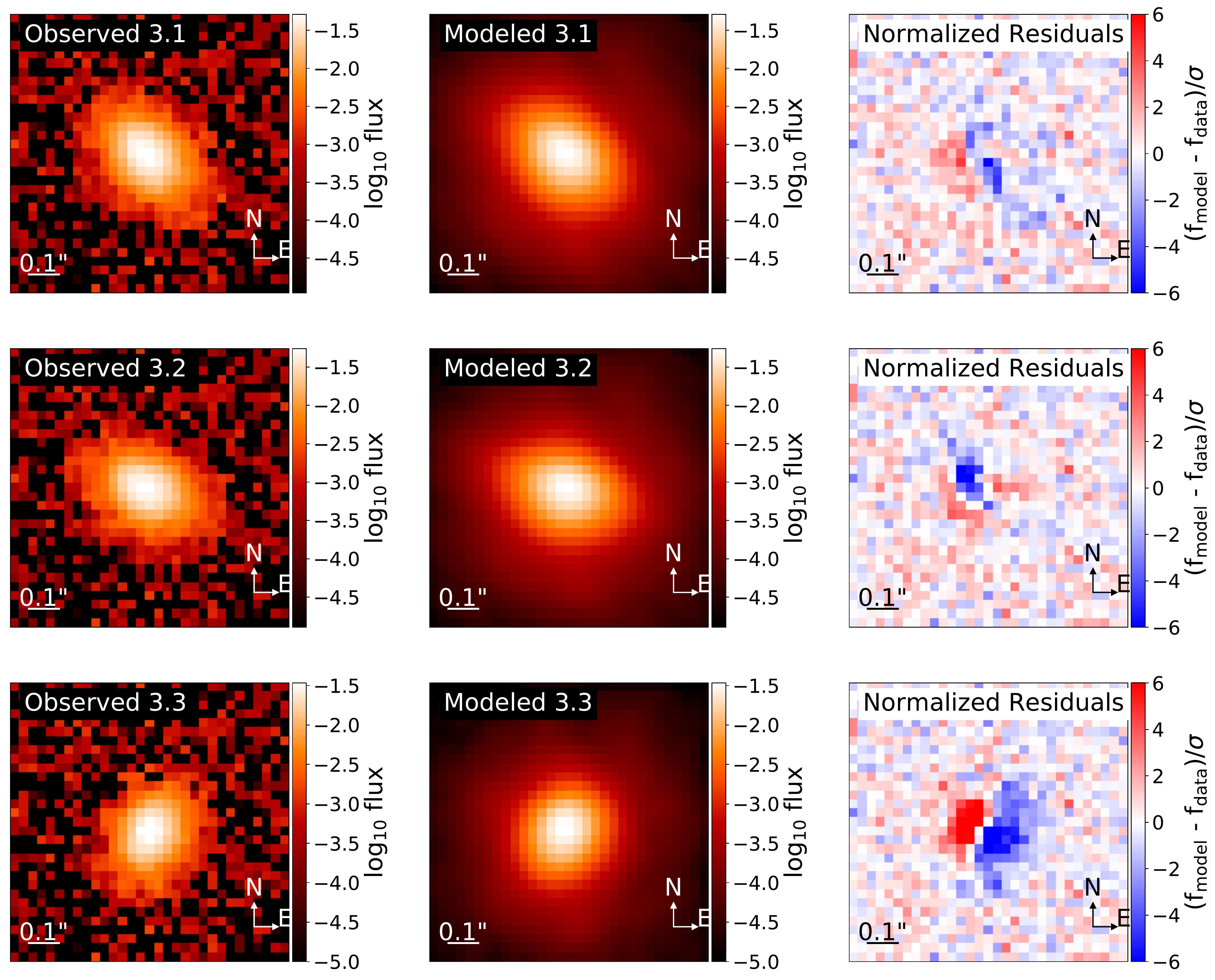}\label{residual3} }\\
 
  \subfloat[Reconstructed source surface brightness distribution of corrected lens models from Brada\v{c}, Williams, CATS, Zitrin-lmt, and Sharon teams, respectively. 
The lens parameters of the least magnified image are fixed (i.e., image 3.3), while of the other images are corrected.
]
{ \includegraphics[width=2\columnwidth]{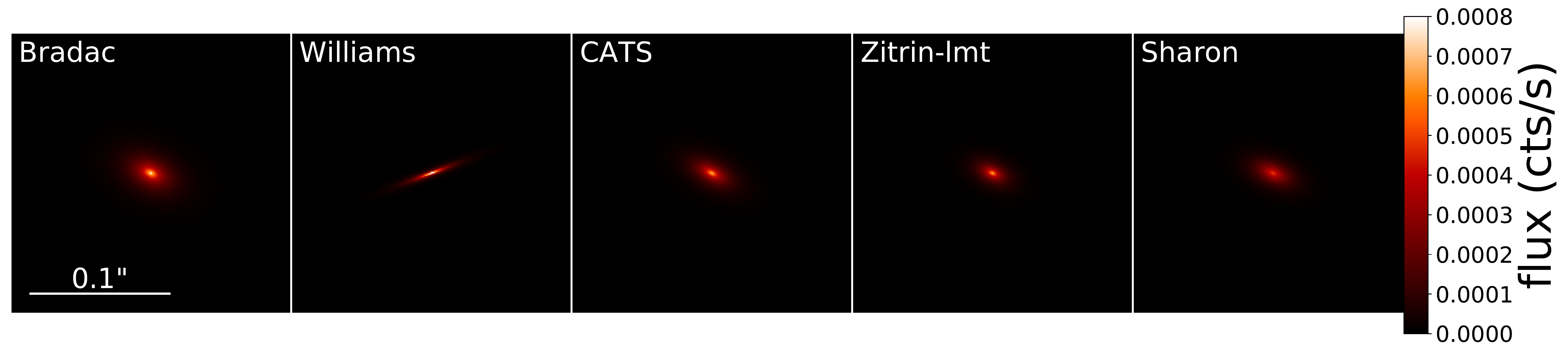} }
  
\caption{Demonstration of the modeling results for multiply-imaged system 3, a) image plane rendition starting from the lens model by the Brada\v{c} team. 
Allowed model complexity includes: shear, convergence acquired from Brada\v{c} team, source model: elliptical S$\'e$rsic. b) reconstructed source surface brightness distribution for different initial models. Allowed model complexity is the same as for the Brada\v{c} model presented in the upper panel.
         }
      \label{fig:recon3}
\end{figure*}

BIC and reduced $\chi^2$ values are recorded in Table \ref{table:comp-sys3}.
The results are shown in Figure~\ref{fig:recon3}.

\begin{figure*}
\centering
\includegraphics[width=2.3\columnwidth]{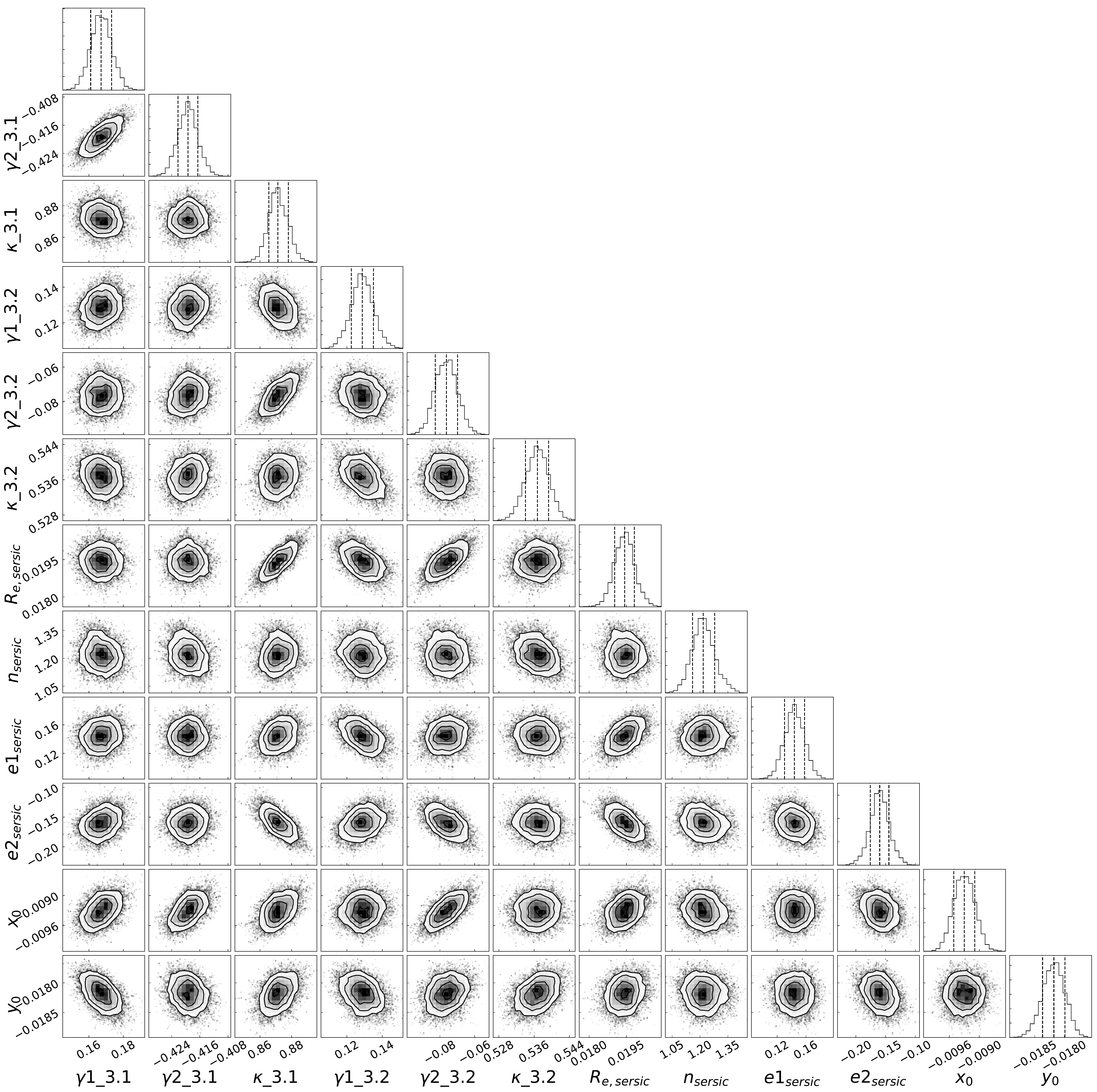}
\caption{MCMC results of multiply-imaged system 3, starting with the Brada$\v c$ lens model. 
Histograms show results of lensing operators $(\gamma1, \gamma2, \kappa)$ for images 3.1 and 3.2. 
Lens parameters of image 3.3 are fixed. The remaining histograms present fitting results of the source light model.
Dashed lines in histograms indicate the uncertainties based on the 16th, 50th, 84th percentiles of the samples.  }
\label{fig:mcmcsys3}
\end{figure*}

MCMC explores the full posterior and provides confidence intervals
as well as degeneracies between parameters.  As an example, we present
the MCMC results of multiply-imaged source 3 with corrections applied to the initial lens model by the Brada\v{c} team in
Figure~\ref{fig:mcmcsys3}.  
The first six histograms show constrained lensing operators of image 3.1 and 3.2 while image 3.3
is fixed.  The uncertainty of the source parameters is shown in the
remaining histograms.

\subsubsection{Modeling details of the multiply-imaged system 4} \label{sec:sys4}

\begin{figure*}
 \centering
    \textbf{Multiply-imaged system 4 with corrected lens parameters }\par\medskip

 \subfloat[Model complexity, lens model: flexion activated and shear, convergence acquired from Brada$\v c$ team, source model: elliptical S$\'e$rsic + $n_{max}=6$.]{\includegraphics[width=1.8\columnwidth]{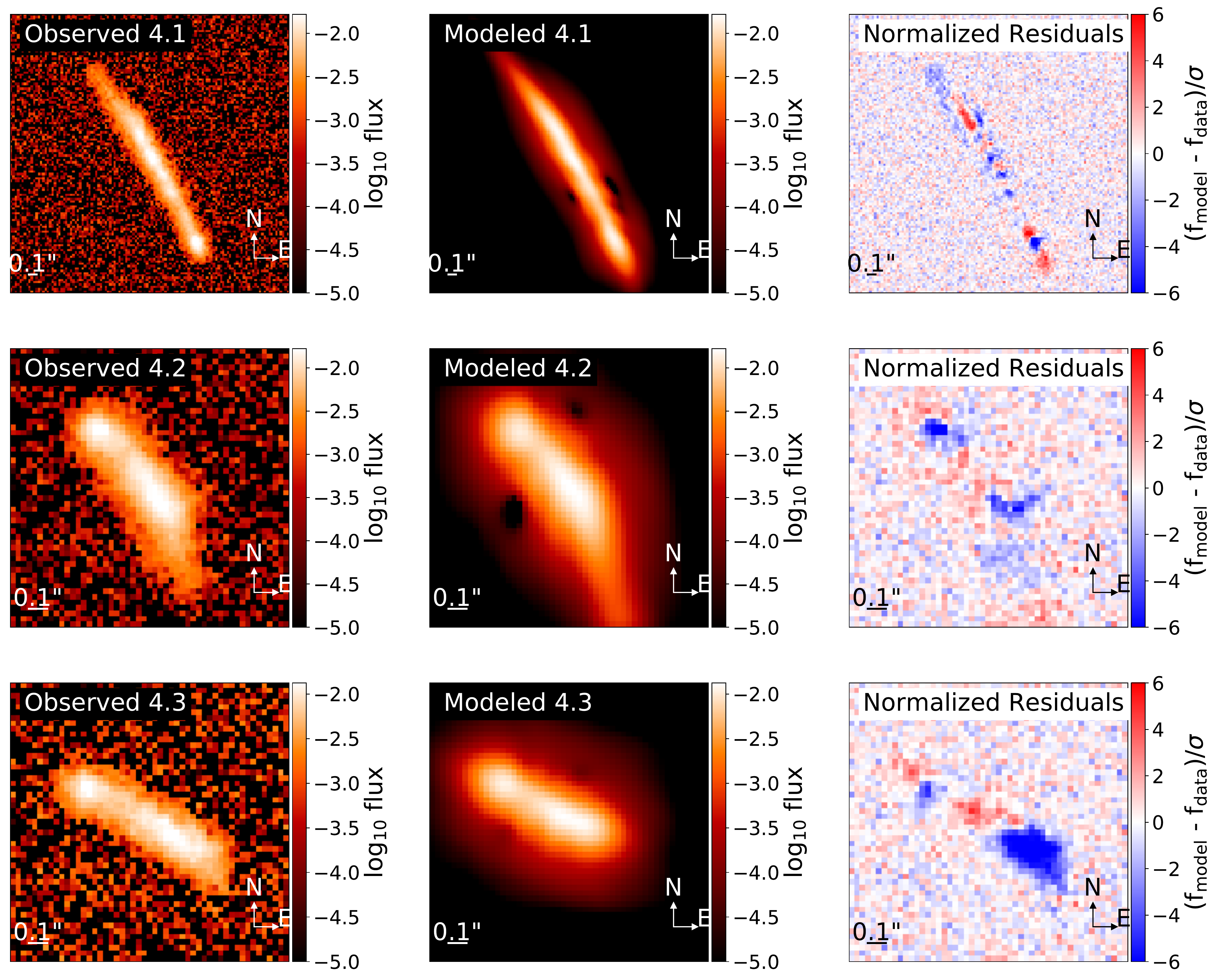} }\\
  \subfloat[Reconstructed source surface brightness distribution via lens models from Brada$\v c$, Williams, CATS, Zitrin-lmt and Sharon teams respectively. ]{ \includegraphics[width=2\columnwidth]{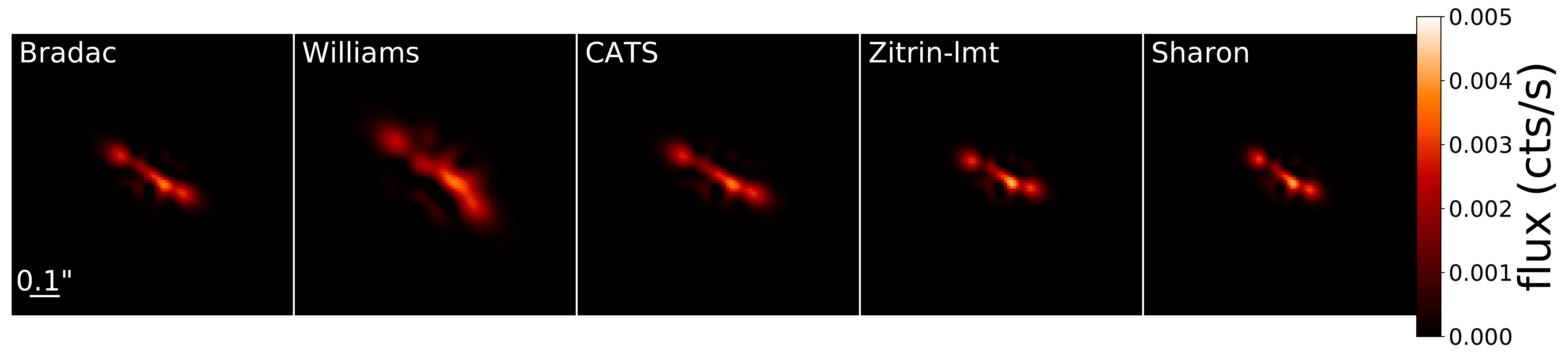} }
  	\caption{Demonstration of the modeling results of the multiply-imaged system 4, same as Figure \ref{fig:recon3}}
	\label{fig:recon4}
\end{figure*}

Images of system 4 are extended and complex, providing more
information to constrain the model. Thus, we can explore the higher model
complexity.  We propose lens model up to flexion and source model with
additional shapelet order $n_{max} = [2, 4, 6]$.  Image 4.3 is
  the least magnified estimated by Brada$\v c$ model, see Table~\ref{table:initial-sys4},
  so we fix its lens parameters. We run a PSO and record the BIC
value for regularization in Table~\ref{table:comp-sys4}.  
As we increase complexity, reduced $\chi^2$ and BIC decrease, indicating that the additional complexity is required.  
Next, we apply additional shapelet bases sequentially. 
The BIC reaches the lowest value $n_{\rm max}=6$ (28 shapelet coefficients). 
We identified the n$_{\rm max}$ range with preliminary tests; BIC does not decrease
significantly beyond 6, indicating that a higher level of source
complexity are not required for this system. The results are shown in
Figure~\ref{fig:recon4}.  Each image in system 4 can also be treated
as a singly-imaged case, as summarized in
Table~\ref{table:initial-sys4}.

\section{Comparison between models}\label{sec:comparison}

Uncertainties in the lens model are a well-known source of systematic
uncertainty in the use of clusters as cosmic telescopes. One of the
goals of \ToolName\ is to allow the wholesale investigation of this
kind of uncertainty in a broad variety of contexts. 
As a primary illustration of the effects of the choice of lens model, that is used to fix the lens parameters of the least magnified image
as well as the improvement introduced by \ToolName\  in the relative distortions,
we compare the reconstructed sources morphologies among different lens models in Section \ref{sec:morph-comp}, and compute how they affect
the source magnitude and effective radius for system 4 in Section
\ref{sec:magsize-comp}. Then, in Section \ref{sec:relative-lens} we
quantify how well the lensing operator is constrained by our procedure
in the case of multiple images.

In this section, the two systems are considered as multiply-imaged, and corrections are applied to the initial lens models.

\subsection{Morphological comparison} \label{sec:morph-comp}

Considering lens models from different teams, we perform similar steps as for the Brada\v{c} model for the two multiply-imaged systems 3 and 4.
The initial magnifications at the position of images among different models span a range of a factor of a few with some prominent outliers. 
For example, the magnifications predicted by the Williams model for system 3 differ up to an order of magnitude from the other ones. 
We also fix the lens parameters of the least magnified image evaluated from the corresponding team, while the others are free. 
For unbiased comparison between the performance of different lens models, we
keep the same range of model complexity as for the Brada\v{c} model. 
Table~\ref{table:comp-sys3} and Table~\ref{table:comp-sys4} show the fitting results of multiply-imaged systems 3 and 4.
While comparing Table~\ref{table:initial-sys3} and Table \ref{table:comp-sys3}, it should be noted that the lens parameters of the least magnified image 3.3 are kept fixed.
While comparing Table~\ref{table:initial-sys4} and Table \ref{table:comp-sys4},  it should be noted that 
Image 4.3 is the fixed one for CATS, Zitrin-lmt, and Sharon teams, while image 4.2 is the least magnified image from the Williams models.
Corrections to the free parameters are expected,  owing to the information supplied by the extended images.
For example, the data show that the magnification ratios are approximately a factor of a few. In contrast, the magnification ratios estimated via initial lens models between images are far larger, underscoring the need for a correction.

We present the comparison of reconstructed sources from
five teams for the two systems in the bottom panels
of Figure~\ref{fig:recon3} and Figure~\ref{fig:recon4}, respectively. 
From left to right, the starting lens models are from Brada$\v c$,
Williams, CATS, Zitrin-lmt, and Sharon teams respectively.  
We clearly see differences in the morphology of the source according to the selected initial lens model. 
In order to quantify the effect on observables, we discuss how the magnitude and half-light radius depend on the estimate of the fixed least magnified image among the different models in Section~\ref{sec:magsize-comp}.

\begin{table*}
	\centering
	\caption{Source properties obtain for systems 3 by considering each image as singly-imaged, i.e. not applying any lensing correction other than that provided by the initial lens models. 
	The first three columns list model names, initial value of lens parameters, and magnification factor of each lensed images. 
	The last two columns show magnitudes F435W AB magnitude (zero-point 25.673)}  and half-light radii of sources reconstructed from image 3.1, 3.2 and 3.3 respectively. 
		\begin{tabular}{lccccccc} 
		\hline
		\hline
Lens Team & $(\gamma1, \gamma2, \kappa)_{3.1}$,  $(\gamma1, \gamma2, \kappa)_{3.2}$,  $(\gamma1, \gamma2, \kappa)_{3.3}$ &  $\mu_{3.1}, \mu_{3.2}, \mu_{3.3}$ & $(m_{3.1}, m_{3.2}, m_{3.3})$ (AB) & $(R_{e3.1}, R_{e3.2}, R_{e3.3}) (")$ \\
\hline
	Brada$\v c$ & (0.43, -0.19, 0.94), (0.15, -0.11, 0.69), (-0.10, 0.22, 0.38) & -4.60, 16.26, 3.07 & (26.60, 28.03, 26.40) & (0.015, 0.008, 0.008) \\
	\hline	
	Williams &    (0.08, 0.19, 0.74), (0.07, 0.25, 0.75), (-0.27, -0.17, 0.58) & 39.84, -204.08, 13.40 & (..., .., 27.88) & (..., ..., 0.006)  \\
	\hline	 
	CATS &  (0.29, -0.21, 0.75), (0.07, -0.13, 0.65), (-0.15, 0.22, 0.47) & -12.53, 9.93, 4.76 & (27.66, 27.55, 26.86) & (0.011, 0.010, 0.007)\\       
	\hline		
	Zitrin-lmt  &  (0.19, -0.27, 0.77), (0.08, -0.19, 0.68), (-0.11, 0.15, 0.58) & -17.83, 16.69, 7.05 &(28.14, 28.07, 27.30)& (0.008, 0.008, 0.005)\\
	\hline	
	Sharon & (0.06, -0.32, 0.94), (0.06, -0.14, 0.76), (-0.23, 0.14, 0.50) & -9.77, 29.07, 5.63 & (27.41, 28.64, 27.07) & (0.010, 0.006, 0.006) \\			        
\hline
	\label{table:initial-sys3}	
	\end{tabular}
\end{table*}

\begin{table*}
	\centering
	\caption{Modeling procedures of multiply-imaged systems 3. Initial lens models from five lens teams, Brada$\v c$, Williams, CATS, Zitrin-lmt and Sharon  respectively. 
	The column Models summarizes the allowed lens and source model complexity.  $\gamma, \kappa$ represent shear, convergence.  ES represents elliptical  S$\'e$rsic. 
	$\chi^2$ and BIC values are recorded in the next columns. The next columns show corrected shear and convergence values of each lensed image. 
	$\mu$ gives magnification evaluated from $\gamma, \kappa$ for each lensed image. The remaining columns list and half-light radius of reconstructed source. }
	\begin{tabular}{lccccccc} 
		\hline
		\hline
Lens Team & Models & $\chi^2$ & BIC &  $(\gamma1, \gamma2, \kappa)_{3.1}$,  $(\gamma1, \gamma2, \kappa)_{3.2}$,  $(\gamma1, \gamma2, \kappa)_{3.3}$ &  $\mu_{3.1}, \mu_{3.2}, \mu_{3.3}$ & $m$ (AB) & $R_e (")$ \\
\hline
Brada$\v c $ & $\gamma, \kappa$, ES & 1.94 &5673.78 & (0.17, -0.42, 0.87), (0.13, -0.08, 0.54), (-0.10, 0.22, 0.38) & -5.31, 5.31, 3.07 & 26.89 & 0.016 \\		        
\hline
Williams & $\gamma, \kappa$, ES & 1.33 & 3935.58 & (-0.17, -0.25, 0.63), (-0.14, -0.21, 0.67), (-0.27, -0.17, 0.58) & 21.98, 22.12, 13.40 & 28.42 & 0.011    \\
\hline	 
CATS &  $\gamma, \kappa$, ES & 1.81 & 5288.12 & (0.11, -0.33, 0.94), (0.07, -0.03, 0.65), (-0.15, 0.22, 0.47) & -8.52, 8.57, 4.76 & 27.39 & 0.013     \\
\hline		
Zitrin-lmt  &  $\gamma, \kappa$, ES & 1.80 & 5237.08 & (0.09, -0.27, 0.93), (0.07, -0.04, 0.71), (-0.11, 0.15, 0.58) & -13.14, 12.89, 7.05 & 27.82 & 0.011 \\
\hline	
Sharon& $\gamma, \kappa$, ES & 1.54 & 4531.69 & (0.09, -0.31, 0.93), (0.06, -0.04, 0.67), (-0.23, 0.14, 0.50) & -10.07, 9.64, 5.63 & 27.57 & 0.012    \\ 
\hline	
\label{table:comp-sys3}	
	\end{tabular}
\end{table*}

\begin{table*}
	\centering
	\caption{Same as the Table \ref{table:initial-sys3} but for images 4.1, 4.2 and 4.3 respectively. }
	\begin{tabular}{lccccccc} 
		\hline
		\hline
Lens Team & $(\gamma1, \gamma2, \kappa)_{4.1}$,  $(\gamma1, \gamma2, \kappa)_{4.2}$,  $(\gamma1, \gamma2, \kappa)_{4.3}$ &  $\mu_{4.1}, \mu_{4.2}, \mu_{4.3}$ & $(m_{4.1}, m_{4.2}, m_{4.3})$ (AB) & $(R_{e4.1}, R_{e4.2}, R_{e4.3}) (")$ \\
\hline
Brada$\v c$ & (-0.02, -0.57, 0.74), (-0.20, -0.15, 0.59), (0.06, -0.02, 0.40) & -3.88, 9.47, 2.81 & (26.83, 27.25, 26.92) & (0.062, 0.062, 0.072) \\
\hline
Williams &   (-0.22, 0.42, 0.88), ( -0.11, -0.02, 0.29), (0.00, 0.06, 0.41)  & -4.75, 2.03, 2.90 &(27.07, 26.54, 27.07) & (0.072, 0.073, 0.072) \\
\hline	
CATS	& ( -0.08 , -0.50 , 0.65), ( -0.12, -0.11, 0.34), (0.03, -0.03, 0.36) & -7.47, 2.46, 2.45   &(26.86, 26.61, 27.04) & (0.067, 0.076, 0.068) \\
 \hline
Zitrin-ltm & (-0.08, -0.39, 0.76), (-0.17, -0.10, 0.55), (0.05, -0.10, 0.43) & -9.91, 6.11, 3.20 &  (27.05, 26.98, 26.90) & (0.061, 0.069, 0.069)\\
 \hline
Sharon &  (-0.14, -0.44, 0.83), (-0.19, -0.09, 0.46), (0.09, -0.08, 0.46), & -5.43, 4.04, 3.61 &(26.77, 26.76, 26.87) & (0.068, 0.067, 0.071)\\
\hline
\label{table:initial-sys4}	
\end{tabular}
\end{table*}

\begin{table*}
	\centering
	\caption{Modeling procedures of multiply-imaged systems 4. Columns are as same as in Table \ref{table:comp-sys3}.	
	$D$ and $n_{max}$ represent flexion $( \mathcal{F}1,\mathcal{F}2,\mathcal{G}1, \mathcal{G}2)$ and shapelets order respectively. }
		\resizebox{\textwidth}{!}{\begin{tabular}{lccccccc} 
		\hline
		\hline
Lens Team & Models & $\chi^2$ & BIC &  $(\gamma1, \gamma2, \kappa)_{4.1}$,  $(\gamma1, \gamma2, \kappa)_{4.2}$,  $(\gamma1, \gamma2, \kappa)_{4.3}$ &  $\mu_{4.1}, \mu_{4.2}, \mu_{4.3}$ & $m$ (AB) & $R_e (")$ \\
\hline	
 Brada$\v c$   & $\gamma, \kappa$, ES & 2.30 & 45615.73 & (-0.11, -0.36, 0.83), (-0.17, 0.06, 0.45),(0.06, -0.02, 0.40) & -8.87, 3.70, 2.81 & 26.11 &0.12 \\ 
 & $\gamma, \kappa, D$, ES & 2.03 & 40440.53 & (-0.10, -0.36, 0.83), (-0.16, 0.08, 0.46),(0.06, -0.02, 0.40) & -9.03, 3.85, 2.81 & 26.10 &0.12 \\ 
 & $\gamma, \kappa, D$, ES, $n_{max}=$2 & 1.65 & 32910.90 & (-0.14, -0.37, 0.73), (-0.15, -0.07, 0.51),(0.06, -0.02, 0.40) & -11.96, 4.70, 2.81 & 26.24 &0.10 \\ 
 & $\gamma, \kappa, D$, ES, $n_{max}=$4 & 1.55 & 31110.69 & (-0.13, -0.38, 0.74), (-0.19, -0.05, 0.48),(0.06, -0.02, 0.40) & -10.67, 4.31, 2.81 & 26.24 &0.11 \\ 
 & $\gamma, \kappa, D$, ES, $n_{max}=$6 & 1.50 & 30175.62 & (-0.12, -0.36, 0.75), (-0.19, -0.04, 0.49),(0.06, -0.02, 0.40) & -12.27, 4.50, 2.81 & 26.30 &0.11 \\ 
\hline

Williams & $\gamma, \kappa$, ES & 2.65 & 52584.85 & (0.28, 0.25, 1.59), (-0.11, -0.02, 0.29),(0.23, -0.13, 0.15) & 4.83, 2.03, 1.53 & 25.48 &0.14 \\ 
 & $\gamma, \kappa, D$, ES & 2.30 & 45642.70 & (0.24, 0.24, 1.57), (-0.11, -0.02, 0.29),(0.19, -0.14, 0.18) & 4.77, 2.03, 1.62 & 25.45 &0.14 \\ 
 & $\gamma, \kappa, D$, ES, $n_{max}=$2 & 1.86 & 37152.67 & (0.24, 0.28, 1.57), (-0.11, -0.02, 0.29),(0.20, -0.17, 0.18) & 5.29, 2.03, 1.66 & 25.49 &0.14 \\ 
 & $\gamma, \kappa, D$, ES, $n_{max}=$4 & 1.77 & 35279.24 & (0.24, 0.28, 1.58), (-0.11, -0.02, 0.29),(0.21, -0.17, 0.17) & 4.99, 2.03, 1.62 & 25.48 &0.14 \\ 
 & $\gamma, \kappa, D$, ES, $n_{max}=$6 & 1.70 & 34153.55 & (0.24, 0.29, 1.59), (-0.11, -0.02, 0.29),(0.20, -0.19, 0.16) & 4.84, 2.03, 1.59 & 25.48 &0.14 \\ 
\hline	
CATS   & $\gamma, \kappa$, ES & 2.30 & 45563.37 & (-0.10, -0.39, 0.82), (-0.23, 0.06, 0.39),(0.03, -0.03, 0.36) & -7.71, 3.17, 2.45 & 25.94 &0.13 \\ 
 & $\gamma, \kappa, D$, ES & 2.02 & 40338.54 & (-0.08, -0.39, 0.84), (-0.26, 0.08, 0.38),(0.03, -0.03, 0.36) & -7.52, 3.22, 2.45 & 25.92 &0.13 \\ 
 & $\gamma, \kappa, D$, ES, $n_{max}=$2 & 1.66 & 33109.41 & (-0.14, -0.39, 0.72), (-0.20, -0.07, 0.46),(0.03, -0.03, 0.36) & -10.72, 4.05, 2.45 & 26.08 &0.11 \\ 
 & $\gamma, \kappa, D$, ES, $n_{max}=$4 & 1.56 & 31312.44 & (-0.13, -0.40, 0.72), (-0.21, -0.06, 0.45),(0.03, -0.03, 0.36) & -10.15, 3.92, 2.45 & 26.10 &0.12 \\ 
 & $\gamma, \kappa, D$, ES, $n_{max}=$6 & 1.51 & 30266.73 & (-0.12, -0.39, 0.74), (-0.22, -0.05, 0.45),(0.03, -0.03, 0.36) & -10.11, 3.97, 2.45 & 26.14 &0.12 \\ 
\hline
Zitrin-ltm & $\gamma, \kappa$, ES & 2.30 & 45608.29 & (-0.11, -0.37, 0.78), (-0.21, -0.02, 0.47),(0.05, -0.10, 0.43) & -9.94, 4.23, 3.20 & 26.23 &0.11 \\ 
 & $\gamma, \kappa, D$, ES & 2.03 & 40490.77 & (-0.10, -0.38, 0.77), (-0.22, -0.00, 0.44),(0.05, -0.10, 0.43) & -9.85, 3.77, 3.20 & 26.14 &0.11 \\ 
 & $\gamma, \kappa, D$, ES, $n_{max}=$2 & 1.65 & 32925.93 & (-0.14, -0.37, 0.71), (-0.17, -0.09, 0.54),(0.05, -0.10, 0.43) & -13.81, 5.73, 3.20 & 26.42 &0.09 \\ 
 & $\gamma, \kappa, D$, ES, $n_{max}=$4 & 1.55 & 31047.41 & (-0.13, -0.39, 0.70), (-0.20, -0.10, 0.49),(0.05, -0.10, 0.43) & -12.66, 4.76, 3.20 & 26.38 &0.09 \\ 
 & $\gamma, \kappa, D$, ES, $n_{max}=$6 & 1.51 & 30287.42 & (-0.12, -0.37, 0.72), (-0.22, -0.10, 0.50),(0.05, -0.10, 0.43) & -13.72, 5.22, 3.20 & 26.46 &0.09 \\ 
 \hline
 Sharon   & $\gamma, \kappa$, ES & 2.30 & 45743.04 & (-0.12, -0.32, 0.81), (-0.13, 0.00, 0.54),(0.09, -0.08, 0.46) & -12.39, 5.14, 3.61 & 26.46 &0.09 \\ 
 & $\gamma, \kappa, D$, ES & 2.05 & 40777.80 & (-0.11, -0.35, 0.80), (-0.17, -0.00, 0.50),(0.09, -0.08, 0.46) & -10.57, 4.52, 3.61 & 26.32 &0.09 \\ 
 & $\gamma, \kappa, D$, ES, $n_{max}=$2 & 1.64 & 32679.63 & (-0.13, -0.34, 0.73), (-0.14, -0.08, 0.57),(0.09, -0.08, 0.46) & -16.78, 6.29, 3.61 & 26.53 &0.08 \\ 
 & $\gamma, \kappa, D$, ES, $n_{max}=$4 & 1.54 & 30757.36 & (-0.12, -0.35, 0.74), (-0.18, -0.06, 0.53),(0.09, -0.08, 0.46) & -14.43, 5.41, 3.61 & 26.53 &0.09 \\ 
 & $\gamma, \kappa, D$, ES, $n_{max}=$6 & 1.50 & 30130.48 & (-0.12, -0.34, 0.74), (-0.19, -0.07, 0.54),(0.09, -0.08, 0.46) & -16.03, 5.86, 3.61 & 26.58 &0.09 \\ 
 
\hline

	\label{table:comp-sys4}	
	\end{tabular} }
\end{table*}

\subsection{Comparison of inferred source magnitude and size}\label{sec:magsize-comp}
\begin{figure*}  
\centering
\includegraphics[width=\columnwidth]{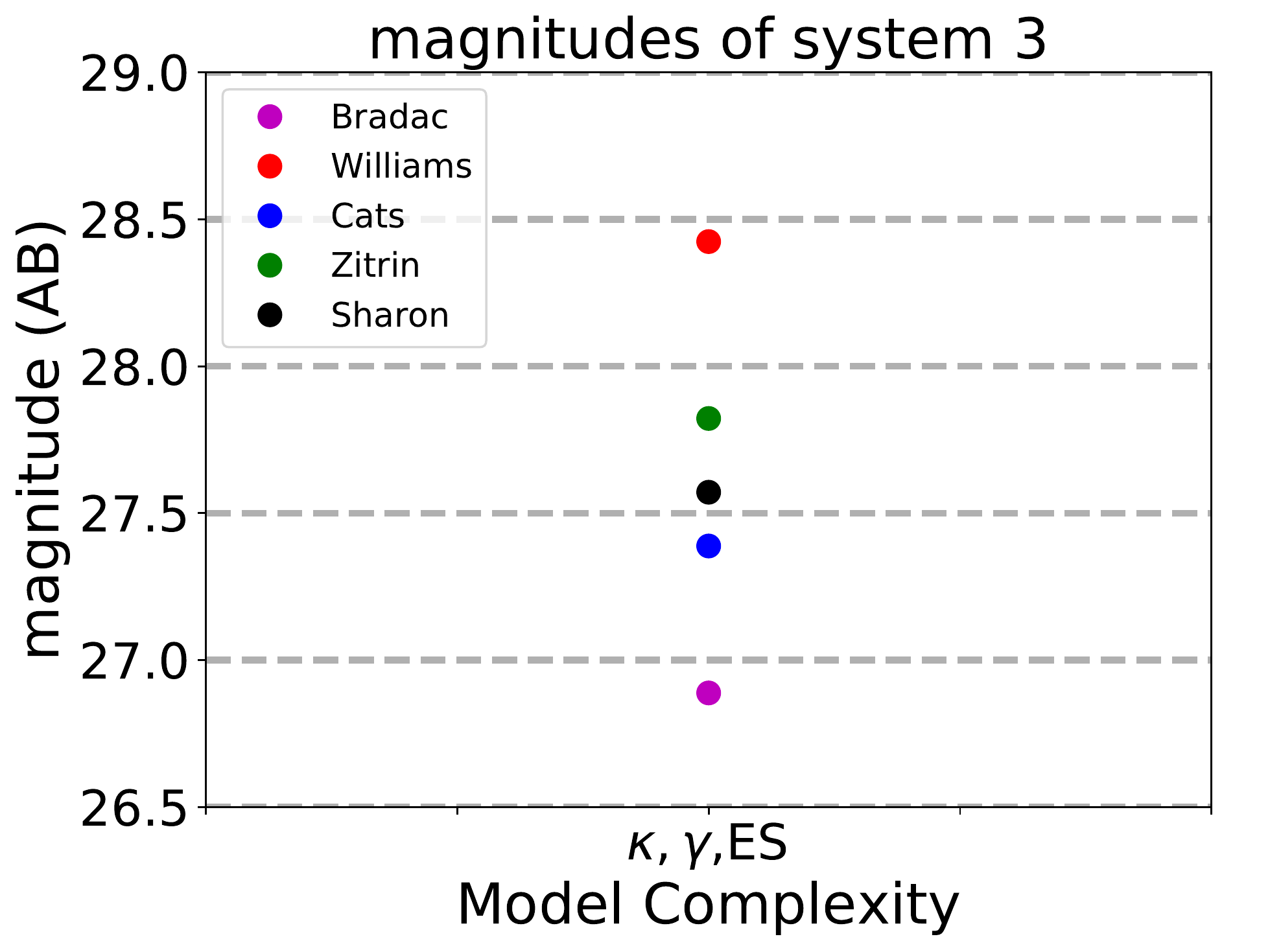} 
\includegraphics[width=\columnwidth]{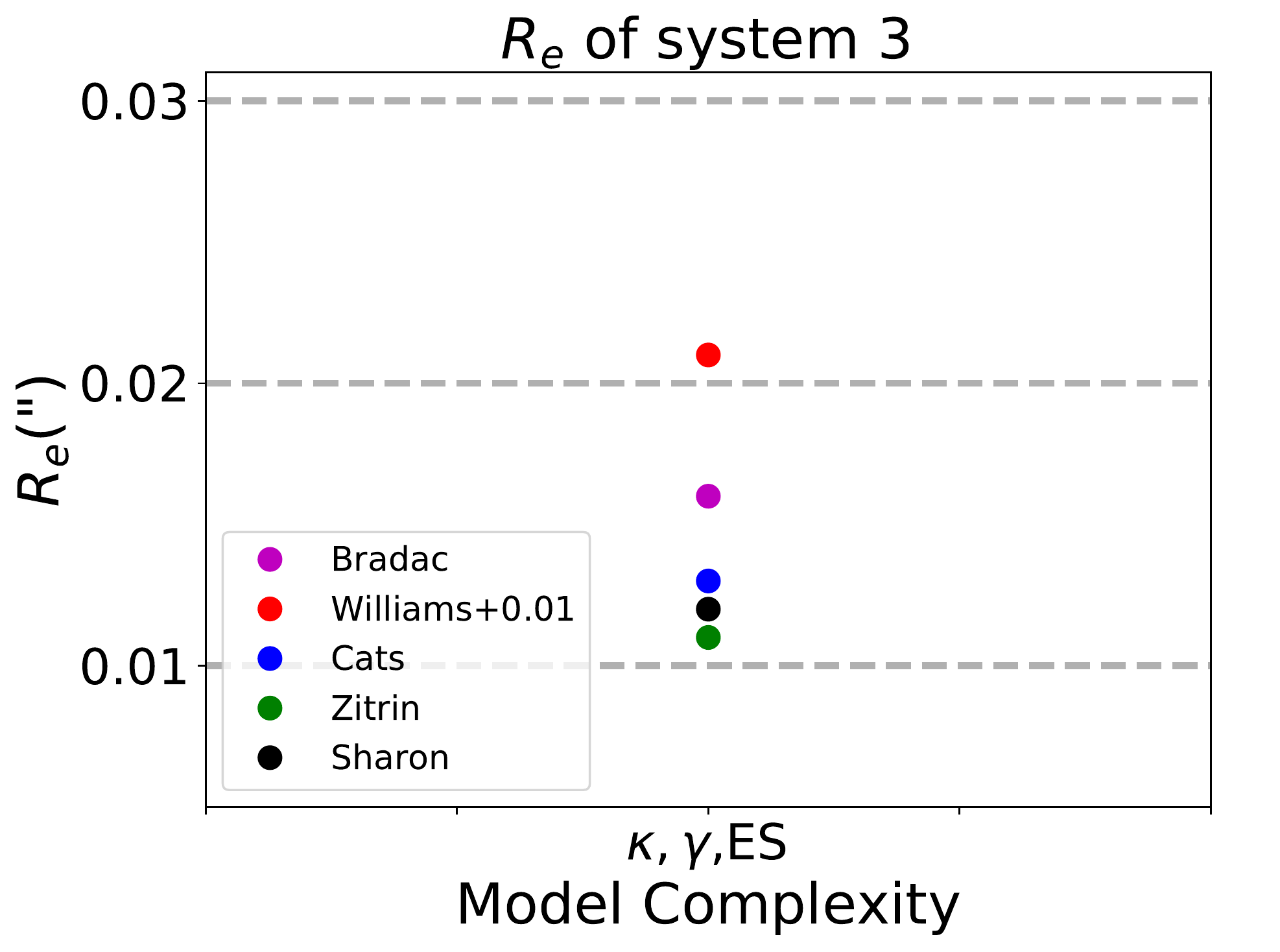}\\
\includegraphics[width=\columnwidth]{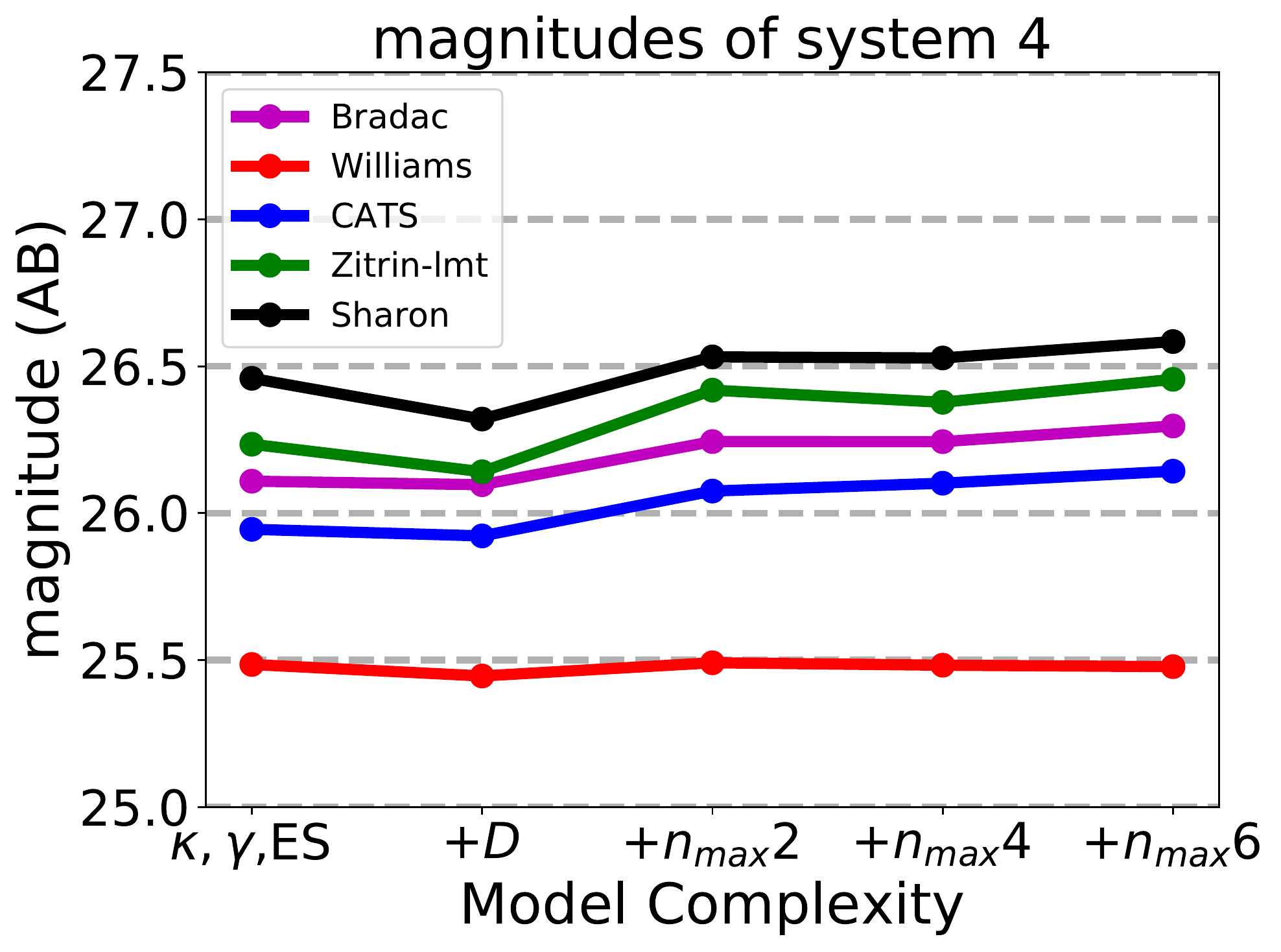} 
\includegraphics[width=\columnwidth]{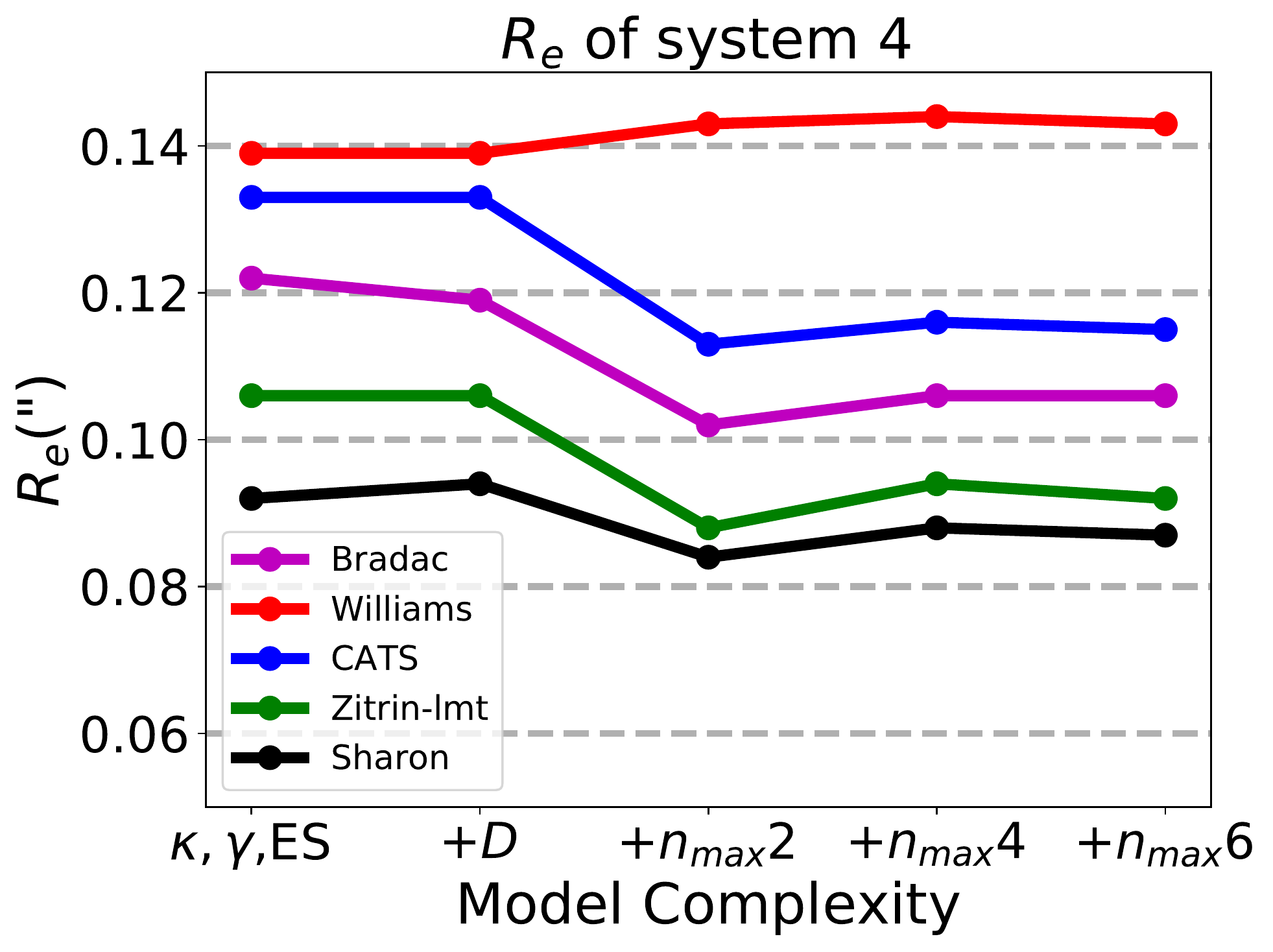}
\caption{Magnitude and half-light radius $R_e$ of reconstructed source for systems 3 and 4, respectively.
Upper) for system 3, the label '$\kappa, \gamma$, ES' indicates the lowest complexity, i.e., shear, convergence and elliptical S$\'e$rsic.
Bottom) for system 4, magnitude and half-light radius as a  function of model complexity. 
The labels '$+D$', '+$n_{max}2$', '+$n_{max}4$', and '+$n_{max}6$' indicate adding flexion and gradually increasing the number of shapelets.  
}
\label{fig:MagRe_sys}
\end{figure*}

The AB Magnitude $m_{AB}$ in source plane is defined in a standard manner as,
\begin{equation}
m_{AB} = -2.5 {\rm log(flux}) + {\rm zeropoint}
\end{equation}
where we adopt as zero-point for ACS F435W 25.673. The half-light radius $R_e$ is defined as the circular aperture that contains half light. 
We present the source plane magnitude and $R_e$ distribution of two systems in  Figure~\ref{fig:MagRe_sys}. Each color represents the result for one of the five lens models.
Numerical values can be found in Table~\ref{table:comp-sys3} and Table~\ref{table:comp-sys4}.
The upper two panels show the distribution of the magnitude and half-light radius $R_e$ for source 3, respectively.
The source magnitude and  $R_e$ span 1.5 magnitudes and 0.01", respectively.
The bottom left panel shows how the magnitude
of source 4 depends on modeling choices. 
During the modeling process, the lens parameters of the least magnified image were kept fixed as initial value acquired from each lens model.
The source magnitude spans a
range of about 1 magnitudes, with r.m.s. scatter 0.39 mags. 
The reconstructed source from the Sharon model is the faintest, while the one
from the Williams model is the brightest. 
Overall, the magnitude is stable for each model, especially after additional model complexity by adding shapelet modes. 
The bottom right panel shows $R_e$ as a function of model complexity. 
The scatter in $R_e$ across models is approximately 18\%, 
while the scatter of the least magnification, i.e., magnifications of the image 4.3, is approximately 25\%.
This dispersion suggests that the differences are fundamentally due to
the mass-sheet degeneracy, and higher-order corrections may be required to improve the agreement between the
models further.

\subsection{Comparison of magnification ratio between lensed images} \label{sec:relative-lens}

\begin{figure*}  
\includegraphics[width=\columnwidth]{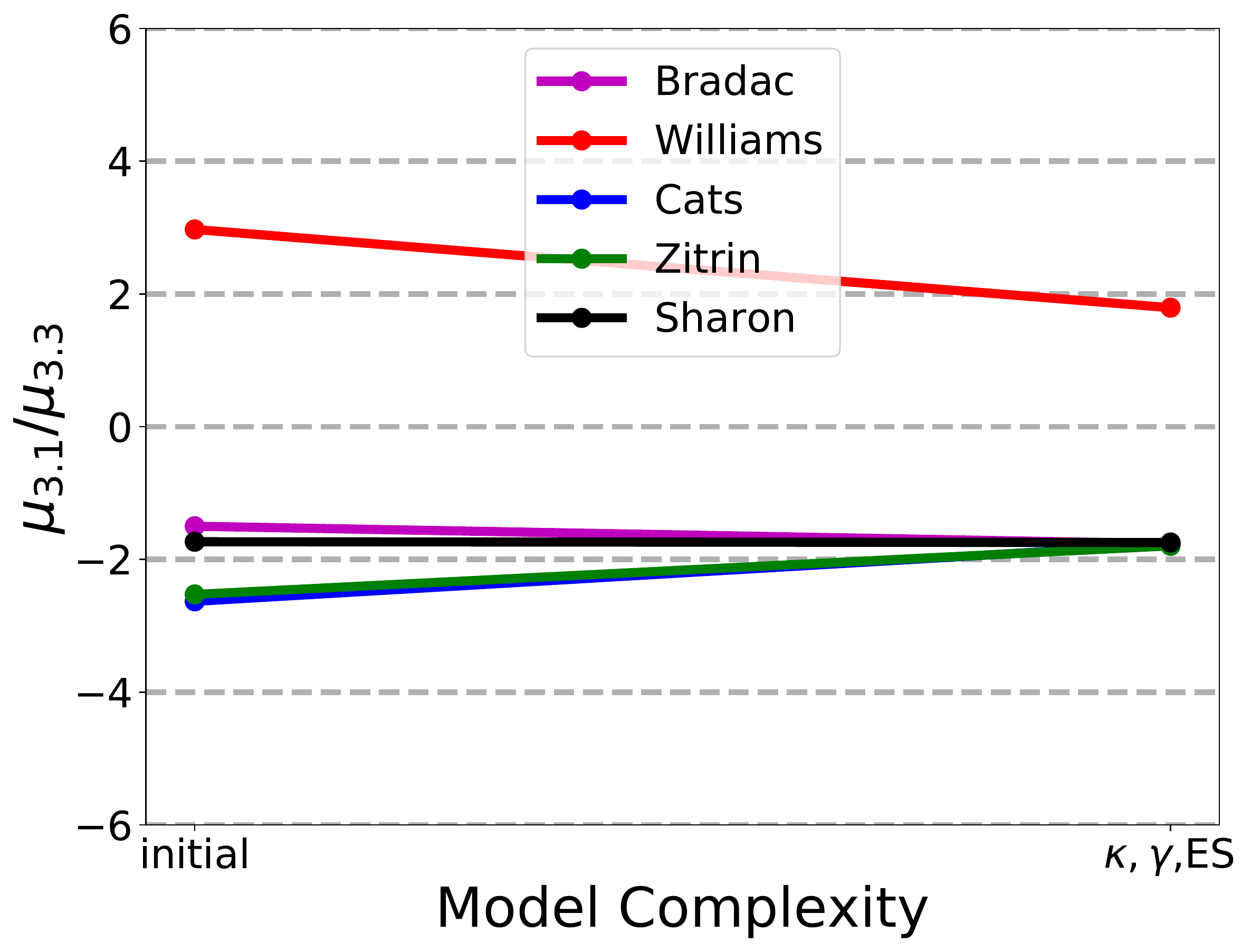}
\includegraphics[width=\columnwidth]{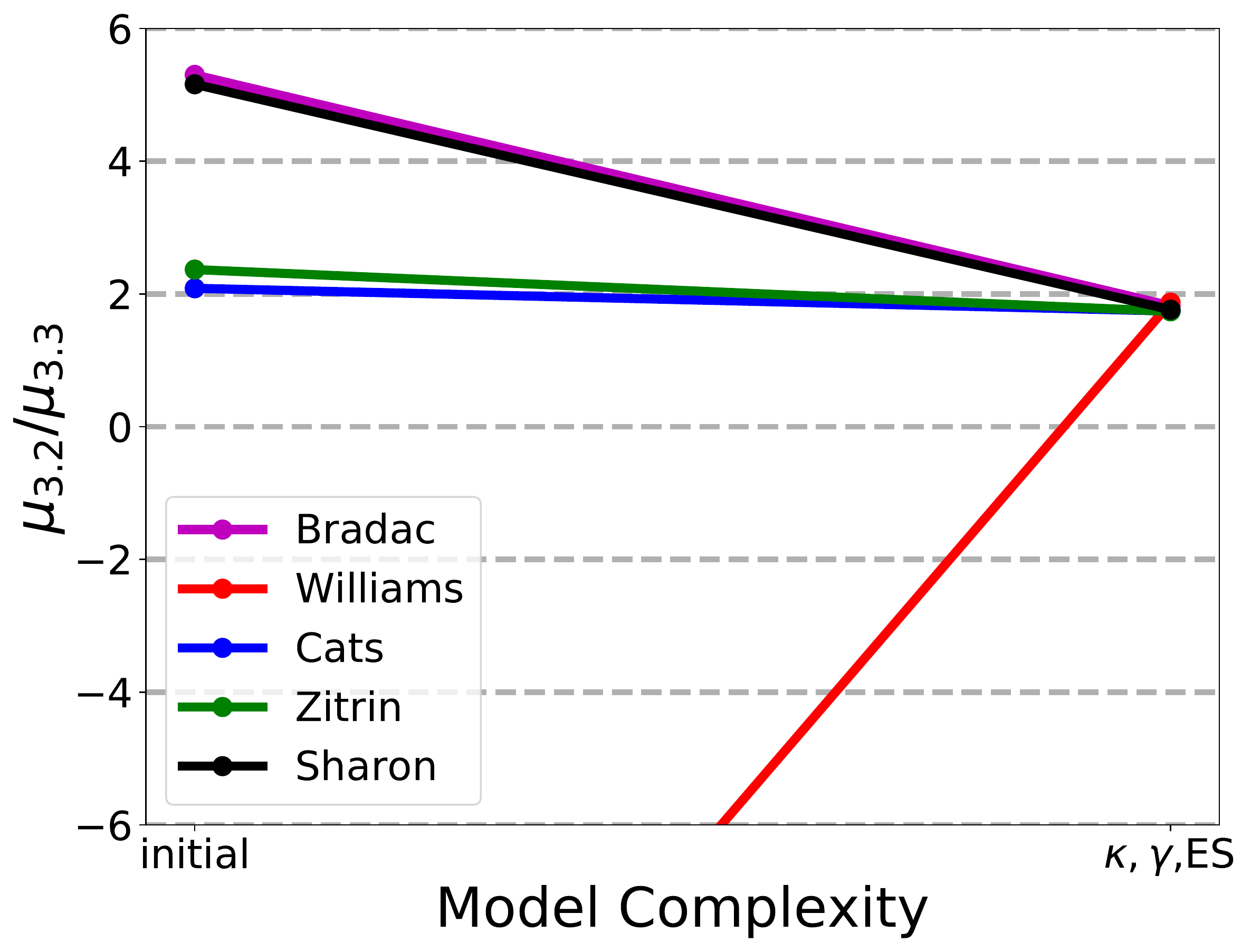} 
\includegraphics[width=\columnwidth]{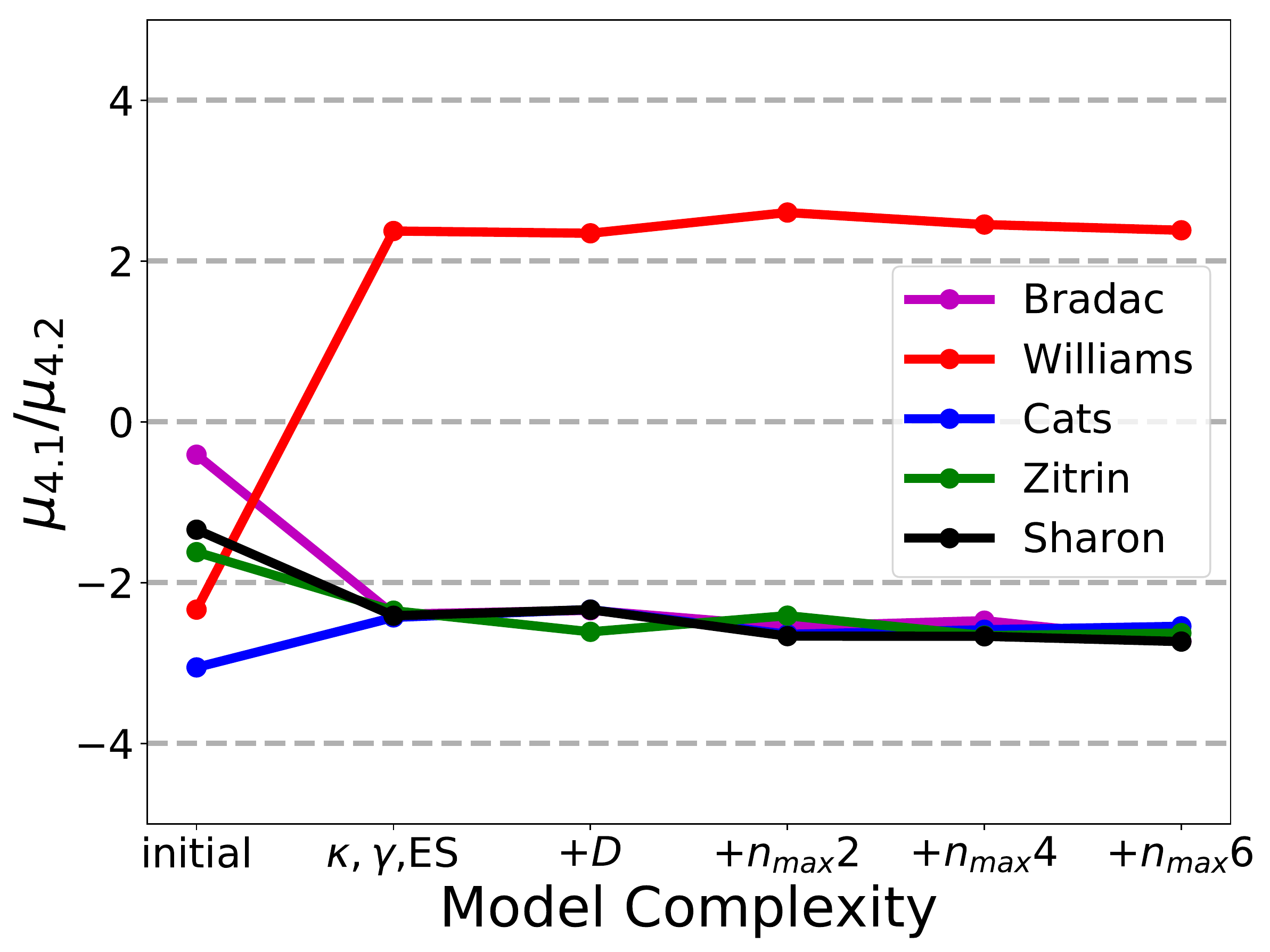}
\includegraphics[width=\columnwidth]{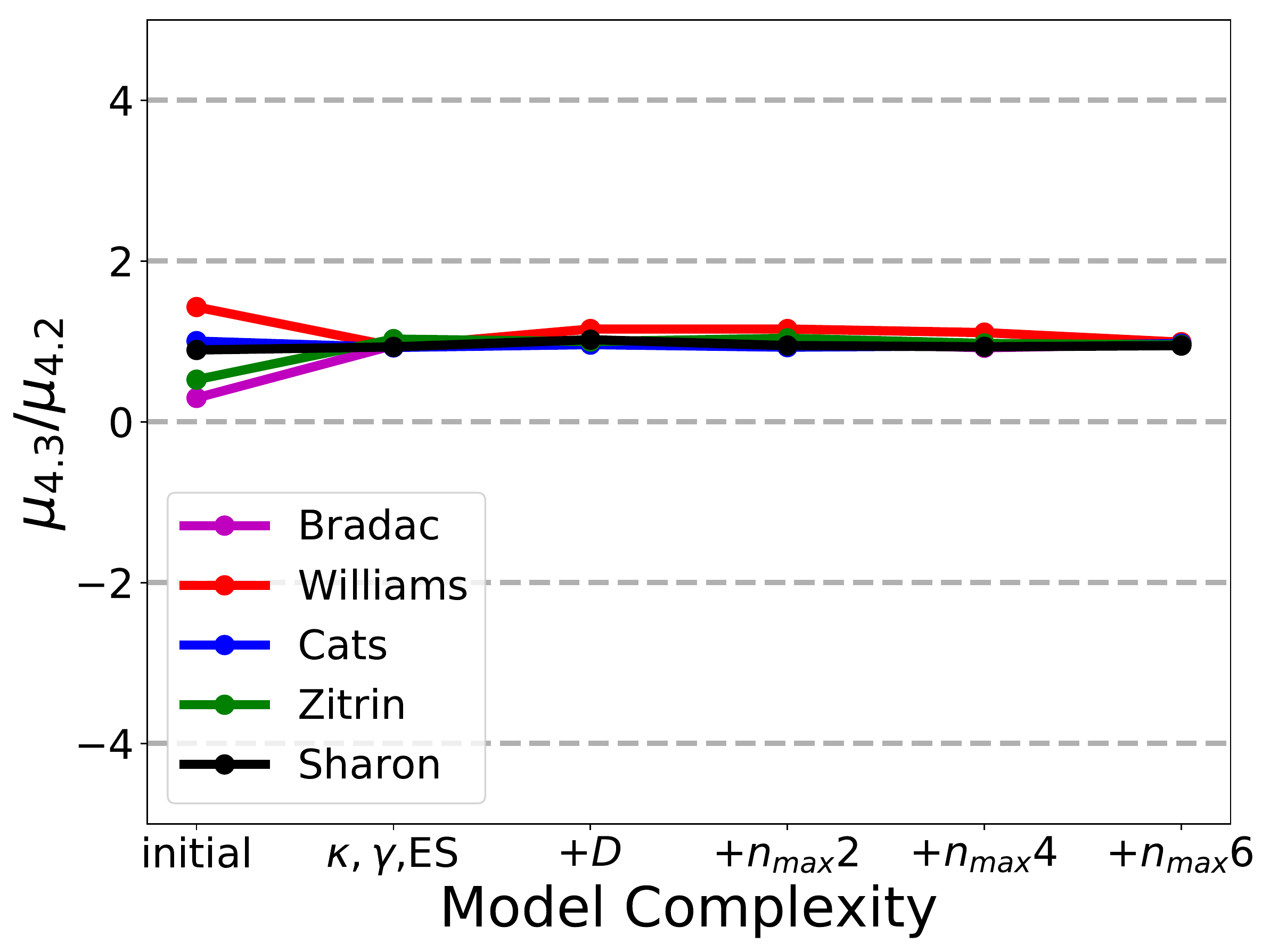}
\caption{Ratio of magnification $\mu$ between two lensed images in two multiply-imaged system respectively. 
Upper left) the ratio between images 3.1 and 3.3. Upper right) ratio between images 3.2 and 3.3. 
Bottom left) the ratio between images 4.1 and 4.2. Bottom right) ratio between images 4.3 and 4.2. 
Label 'initial' indicates value acquired directly from the lens model, others are the same as Figure \ref{fig:MagRe_sys}. 
Colors represent five lens models. }
\label{fig:relative_sys34}
\end{figure*}

The relative magnification of the images is expected to be tightly constrained by the data.
We show relative magnifications $\mu$ as a function of model complexity for each initial lens model in Figure~\ref{fig:relative_sys34}, color coded as in Figure~\ref{fig:MagRe_sys}.
It is clear that the initial estimates of $\mu$ differ dramatically for each lens model (see also Table~\ref{table:initial-sys3}.)
For example, $\mu$ of image 3.2 varies from $\sim -204.08$ to $\sim 9.93$.
The discrepancy is not as dramatic for the absolute value of the relative magnifications, although it is still quite substantial. 

After applying corrections to the initial lens models, we observe a
definite improvement in the consistency of the magnification ratio. 
In Figure~\ref{fig:relative_sys34}, the absolute values of relative $\mu_{4.1}/\mu_{4.2}$ and
$\mu_{3.1}/\mu_{3.3}$ get close to $\sim2$.  In those two cases, the
parity is not well determined (hence the plus/minus dichotomy) because
the image does not have much structure perpendicular to the highly
stretched direction.  The ratios $\mu_{4.3}/\mu_{4.2}$ and
$\mu_{3.2}/\mu_{3.3}$ are well constrained to $\sim1$ and $\sim2$, even though the initial models had substantial scatter.

\section{Summary and Conclusion} \label{sum-conclusion}

From the perspective of lens modeling, cluster-scale
lensing is full of challenges. Most state-of-the-art models are
constrained exclusively by the positions of the lensed images rather than via extended source reconstruction.  
Therefore, additional efforts are needed to reconstruct the background source and determine
the uncertainties associated with the lens model. 
Manually, performing source reconstruction on a handful of data is feasible. 
However, with
current and upcoming observations, the quality and quantity of
cluster lensing data are expected to improve dramatically.

In order to prepare for such an explosion of data, we have developed
and made public a fast and versatile tool \ToolName\ . It adopts
a forward modeling approach to perform source reconstruction with
corrections on the initial lens parameters, taking into account the
blurring of the PSF. \ToolName\ is implemented in python, building on
the publicly available code \textsc{lenstronomy}. 

In this paper, we describe the current implementation of \ToolName\ as
well as present the first illustration of its capabilities using two
sets of multiple images in the HFF cluster MACS0717+3745, starting
from five publicly available models as the initial guess of the
lensing potential. One system is chosen to be of a compact source,
providing limited information to correct the lens model. The other
system is significantly extended and enables exploration of
complexity, including corrections up to flexion order in the lens model
and up to S\'ersic + shapelets (with $n_{\rm max}$ up to 6) in the source model.

We find that the choice of initial lens model with the lens model at the least magnified image held fixed affects the inferred
source magnitude and size at the level of 0.39 mag and 18\%
r.m.s. scatter respectively.
The scatter does not reduce by increasing the complexity of corrections on the lens model within the range considered in this work. 

In contrast, the absolute ratio of magnifications between the images
converges rapidly to a common value, even though the initial lens
models provided sometimes dramatically different estimates.

In conclusion, we observe that our correction scheme produces a
substantial improvement in the relative magnification, i.e., the
quantity directly constrained by the data. Starting with significantly
different initial models, the correction schemes makes them all
converge to very similar absolute magnification ratios. 
The convergence is significantly more pronounced than observation in 
the case of magnitudes and effective radii, which are absolute
quantities, and thus their measurement depends on breaking the mass
sheet degeneracy.

In the future, we plan to carry out a similar investigation for a large
sample of objects. The ultimate goal is measuring the irreducible
scatter of current state of the art lens models as a way to quantify
this source of systematic uncertainty in the estimation of the size
luminosity/mass relation, and other observables, through cosmic
telescopes (Yang et al. 2020, in preparation).
 
\section*{Acknowledgements}
This work utilizes gravitational lensing models produced by PIs Brada\v{c}, Natarajan \& Kneib (CATS), Merten \& Zitrin, Sharon, Williams, Keeton, Bernstein and Diego, and the GLAFIC group. This lens modeling was partially funded by the HST Frontier Fields program conducted by STScI. STScI is operated by the Association of Universities for Research in Astronomy, Inc. under NASA contract NAS 5-26555. The lens models were obtained from the Mikulski Archive for Space Telescopes (MAST). LY, SB and TT acknowledge support from NASA through grant HST-GO-14630. LY is supported from the China Scholarship Council. TT acknowledges support by NASA through grant number JWST-ERS-01324.001 " Through the looking GLASS: a JWST exploration of Galaxy Formation and Evolution from Cosmic Dawn to Present Day'' from the Space Telescope Science Institute, which is operated by AURA, Inc., under NASA contract NAS 5-26555. The authors thank Maru\v{s}a Brada\v{c}, Xuheng Ding,  Austin Hoag,  Anowar J. Shajib, Guido Roberts-Borsani, and Jenny Wagner for several discussions that helped shaped the code and manuscript.



\bibliographystyle{mnras}
\bibliography{clusterlens} 

\begin{thebibliography}{}
\makeatletter
\relax
\def\mn@urlcharsother{\let\do\@makeother \do\$\do\&\do\#\do\^\do\_\do\%\do\~}
\def\mn@doi{\begingroup\mn@urlcharsother \@ifnextchar [ {\mn@doi@}
  {\mn@doi@[]}}
\def\mn@doi@[#1]#2{\def\@tempa{#1}\ifx\@tempa\@empty \href
  {http://dx.doi.org/#2} {doi:#2}\else \href {http://dx.doi.org/#2} {#1}\fi
  \endgroup}
\def\mn@eprint#1#2{\mn@eprint@#1:#2::\@nil}
\def\mn@eprint@arXiv#1{\href {http://arxiv.org/abs/#1} {{\tt arXiv:#1}}}
\def\mn@eprint@dblp#1{\href {http://dblp.uni-trier.de/rec/bibtex/#1.xml}
  {dblp:#1}}
\def\mn@eprint@#1:#2:#3:#4\@nil{\def\@tempa {#1}\def\@tempb {#2}\def\@tempc
  {#3}\ifx \@tempc \@empty \let \@tempc \@tempb \let \@tempb \@tempa \fi \ifx
  \@tempb \@empty \def\@tempb {arXiv}\fi \@ifundefined
  {mn@eprint@\@tempb}{\@tempb:\@tempc}{\expandafter \expandafter \csname
  mn@eprint@\@tempb\endcsname \expandafter{\@tempc}}}

\bibitem[\protect\citeauthoryear{Acebron, Jullo, Limousin, Tilquin, Giocoli,
  Jauzac, Mahler  \& Richard}{Acebron et~al.}{2017}]{Acebron2017}
Acebron A.,  Jullo E.,  Limousin M.,  Tilquin A.,  Giocoli C.,  Jauzac M.,
  Mahler G.,   Richard J.,  2017, \mn@doi [Monthly Notices of the Royal
  Astronomical Society] {10.1093/mnras/stx1330}, 470, 1809

\bibitem[\protect\citeauthoryear{{Birrer} \& {Amara}}{{Birrer} \&
  {Amara}}{2018}]{Birrer&Amara2018}
{Birrer} S.,  {Amara} A.,  2018, \mn@doi [Physics of the Dark Universe]
  {10.1016/j.dark.2018.11.002}, \href
  {https://ui.adsabs.harvard.edu/abs/2018PDU....22..189B} {22, 189}

\bibitem[\protect\citeauthoryear{{Birrer}, {Amara}  \& {Refregier}}{{Birrer}
  et~al.}{2015}]{Birrer2015}
{Birrer} S.,  {Amara} A.,   {Refregier} A.,  2015, \mn@doi [\apj]
  {10.1088/0004-637X/813/2/102}, \href
  {https://ui.adsabs.harvard.edu/abs/2015ApJ...813..102B} {813, 102}

\bibitem[\protect\citeauthoryear{{Birrer}, {Amara}  \& {Refregier}}{{Birrer}
  et~al.}{2016}]{Birrer:2016}
{Birrer} S.,  {Amara} A.,   {Refregier} A.,  2016, \mn@doi [\jcap]
  {10.1088/1475-7516/2016/08/020}, \href
  {https://ui.adsabs.harvard.edu/abs/2016JCAP...08..020B} {2016, 020}

\bibitem[\protect\citeauthoryear{{Birrer}, {Amara}  \& {Refregier}}{{Birrer}
  et~al.}{2017}]{Birrer:2017substructure}
{Birrer} S.,  {Amara} A.,   {Refregier} A.,  2017, \mn@doi [\jcap]
  {10.1088/1475-7516/2017/05/037}, \href
  {https://ui.adsabs.harvard.edu/abs/2017JCAP...05..037B} {2017, 037}

\bibitem[\protect\citeauthoryear{{Birrer} et~al.,}{{Birrer}
  et~al.}{2019}]{Birrer2019}
{Birrer} S.,  et~al., 2019, \mn@doi [\mnras] {10.1093/mnras/stz200}, \href
  {https://ui.adsabs.harvard.edu/abs/2019MNRAS.484.4726B} {484, 4726}

\bibitem[\protect\citeauthoryear{{Blandford}, {Surpi}  \&
  {Kundi{\'c}}}{{Blandford} et~al.}{2001}]{Blandford2001}
{Blandford} R.,  {Surpi} G.,   {Kundi{\'c}} T.,  2001, in {Brainerd} T.~G.,
  {Kochanek} C.~S.,  eds,  Astronomical Society of the Pacific Conference
  Series Vol. 237, Gravitational Lensing: Recent Progress and Future Go. p.~65
  (\mn@eprint {} {astro-ph/0001496})

\bibitem[\protect\citeauthoryear{Blandford, Suyu, Marshall  \&
  Hobson}{Blandford et~al.}{2006}]{Suyu2006}
Blandford R.~D.,  Suyu S.~H.,  Marshall P.~J.,   Hobson M.~P.,  2006, \mn@doi
  [Monthly Notices of the Royal Astronomical Society]
  {10.1111/j.1365-2966.2006.10733.x}, 371, 983

\bibitem[\protect\citeauthoryear{{Bouwens}, {Illingworth}, {Oesch}, {Atek},
  {Lam}  \& {Stefanon}}{{Bouwens} et~al.}{2017}]{Bouwens2017}
{Bouwens} R.~J.,  {Illingworth} G.~D.,  {Oesch} P.~A.,  {Atek} H.,  {Lam} D.,
  {Stefanon} M.,  2017, \mn@doi [\apj] {10.3847/1538-4357/aa74e4}, \href
  {https://ui.adsabs.harvard.edu/abs/2017ApJ...843...41B} {843, 41}

\bibitem[\protect\citeauthoryear{{Brada{\v c}}, {Schneider}, {Lombardi}  \&
  {Erben}}{{Brada{\v c}} et~al.}{2005}]{Bradac2005}
{Brada{\v c}} M.,  {Schneider} P.,  {Lombardi} M.,   {Erben} T.,  2005, \mn@doi
  [\aap] {10.1051/0004-6361:20042233}, \href
  {https://ui.adsabs.harvard.edu/abs/2005A%26A...437...39B} {437, 39}

\bibitem[\protect\citeauthoryear{{Brada{\v c}} et~al.,}{{Brada{\v c}}
  et~al.}{2009}]{Bradac2009}
{Brada{\v c}} M.,  et~al., 2009, \mn@doi [\apj] {10.1088/0004-637X/706/2/1201},
  \href {https://ui.adsabs.harvard.edu/abs/2009ApJ...706.1201B} {706, 1201}

\bibitem[\protect\citeauthoryear{{Brada{\v{c}}}, {Lombardi}  \&
  {Schneider}}{{Brada{\v{c}}} et~al.}{2004}]{bradaclomsch2004}
{Brada{\v{c}}} M.,  {Lombardi} M.,   {Schneider} P.,  2004, \mn@doi [\aap]
  {10.1051/0004-6361:20035744}, \href
  {https://ui.adsabs.harvard.edu/abs/2004A&A...424...13B} {424, 13}

\bibitem[\protect\citeauthoryear{Bradley et~al.,}{Bradley
  et~al.}{2019}]{Bradley2019}
Bradley L.,  et~al., 2019, astropy/photutils: v0.7,
  \mn@doi{10.5281/zenodo.2533376}, \url
  {https://doi.org/10.5281/zenodo.2533376}

\bibitem[\protect\citeauthoryear{{Caminha} et~al.,}{{Caminha}
  et~al.}{2017}]{Caminha2017}
{Caminha} G.~B.,  et~al., 2017, \mn@doi [\aap] {10.1051/0004-6361/201731498},
  \href {https://ui.adsabs.harvard.edu/abs/2017A&A...607A..93C} {607, A93}

\bibitem[\protect\citeauthoryear{{Diaz Rivero} \& {Dvorkin}}{{Diaz Rivero} \&
  {Dvorkin}}{2019}]{DiazRivero:2019}
{Diaz Rivero} A.,  {Dvorkin} C.,  2019, arXiv e-prints, \href
  {https://ui.adsabs.harvard.edu/abs/2019arXiv191000015D} {p. arXiv:1910.00015}

\bibitem[\protect\citeauthoryear{{Ding} et~al.,}{{Ding}
  et~al.}{2019}]{Ding2019}
{Ding} X.,  et~al., 2019, arXiv e-prints, \href
  {https://ui.adsabs.harvard.edu/abs/2019arXiv191011875D} {p. arXiv:1910.11875}

\bibitem[\protect\citeauthoryear{{Ebeling}, {Ma}  \& {Barrett}}{{Ebeling}
  et~al.}{2014}]{Ebeling2014}
{Ebeling} H.,  {Ma} C.-J.,   {Barrett} E.,  2014, \mn@doi [\apjs]
  {10.1088/0067-0049/211/2/21}, \href
  {https://ui.adsabs.harvard.edu/abs/2014ApJS..211...21E} {211, 21}

\bibitem[\protect\citeauthoryear{{Falco}, {Gorenstein}  \& {Shapiro}}{{Falco}
  et~al.}{1985}]{Falco:1985}
{Falco} E.~E.,  {Gorenstein} M.~V.,   {Shapiro} I.~I.,  1985, \mn@doi [\apjl]
  {10.1086/184422}, \href
  {https://ui.adsabs.harvard.edu/abs/1985ApJ...289L...1F} {289, L1}

\bibitem[\protect\citeauthoryear{{Foreman-Mackey}, {Hogg}, {Lang}  \&
  {Goodman}}{{Foreman-Mackey} et~al.}{2013}]{Foreman-Mackey2013}
{Foreman-Mackey} D.,  {Hogg} D.~W.,  {Lang} D.,   {Goodman} J.,  2013, \mn@doi
  [\pasp] {10.1086/670067}, \href
  {https://ui.adsabs.harvard.edu/abs/2013PASP..125..306F} {125, 306}

\bibitem[\protect\citeauthoryear{Gilman, Birrer, Treu, Nierenberg  \&
  Benson}{Gilman et~al.}{2019}]{Gilman2019}
Gilman D.,  Birrer S.,  Treu T.,  Nierenberg A.,   Benson A.,  2019, \mn@doi
  [Monthly Notices of the Royal Astronomical Society] {10.1093/mnras/stz1593},
  487, 5721

\bibitem[\protect\citeauthoryear{{Gorenstein}, {Falco}  \&
  {Shapiro}}{{Gorenstein} et~al.}{1988}]{Gorenstein1988}
{Gorenstein} M.~V.,  {Falco} E.~E.,   {Shapiro} I.~I.,  1988, \mn@doi [\apj]
  {10.1086/166226}, \href
  {https://ui.adsabs.harvard.edu/abs/1988ApJ...327..693G} {327, 693}

\bibitem[\protect\citeauthoryear{{Grillo} et~al.,}{{Grillo}
  et~al.}{2018}]{Grillo2018}
{Grillo} C.,  et~al., 2018, \mn@doi [\apj] {10.3847/1538-4357/aac2c9}, \href
  {https://ui.adsabs.harvard.edu/abs/2018ApJ...860...94G} {860, 94}

\bibitem[\protect\citeauthoryear{Hoekstra, Bartelmann, Dahle, Israel, Limousin
  \& Meneghetti}{Hoekstra et~al.}{2013}]{Hoekstra2013}
Hoekstra H.,  Bartelmann M.,  Dahle H.,  Israel H.,  Limousin M.,   Meneghetti
  M.,  2013, \mn@doi [Space Science Reviews] {10.1007/s11214-013-9978-5}, 177,
  75

\bibitem[\protect\citeauthoryear{{Johnson}, {Sharon}, {Bayliss}, {Gladders},
  {Coe}  \& {Ebeling}}{{Johnson} et~al.}{2014}]{Sharon2014Johnson}
{Johnson} T.~L.,  {Sharon} K.,  {Bayliss} M.~B.,  {Gladders} M.~D.,  {Coe} D.,
   {Ebeling} H.,  2014, \mn@doi [\apj] {10.1088/0004-637X/797/1/48}, \href
  {https://ui.adsabs.harvard.edu/abs/2014ApJ...797...48J} {797, 48}

\bibitem[\protect\citeauthoryear{{Jones} et~al.,}{{Jones}
  et~al.}{2015}]{Jones2015}
{Jones} T.,  et~al., 2015, \mn@doi [\aj] {10.1088/0004-6256/149/3/107}, \href
  {https://ui.adsabs.harvard.edu/abs/2015AJ....149..107J} {149, 107}

\bibitem[\protect\citeauthoryear{{Joseph}, {Courbin}, {Starck}  \&
  {Birrer}}{{Joseph} et~al.}{2019}]{Joseph2019}
{Joseph} R.,  {Courbin} F.,  {Starck} J.~L.,   {Birrer} S.,  2019, \mn@doi
  [\aap] {10.1051/0004-6361/201731042}, \href
  {https://ui.adsabs.harvard.edu/abs/2019A&A...623A..14J} {623, A14}

\bibitem[\protect\citeauthoryear{{Jullo}, {Natarajan}, {Kneib}, {D'Aloisio},
  {Limousin}, {Richard}  \& {Schimd}}{{Jullo} et~al.}{2010}]{Jullo2010}
{Jullo} E.,  {Natarajan} P.,  {Kneib} J.~P.,  {D'Aloisio} A.,  {Limousin} M.,
  {Richard} J.,   {Schimd} C.,  2010, \mn@doi [Science]
  {10.1126/science.1185759}, \href
  {https://ui.adsabs.harvard.edu/abs/2010Sci...329..924J} {329, 924}

\bibitem[\protect\citeauthoryear{{Kawamata}, {Ishigaki}, {Shimasaku}, {Oguri},
  {Ouchi}  \& {Tanigawa}}{{Kawamata} et~al.}{2018}]{Kawamata2018}
{Kawamata} R.,  {Ishigaki} M.,  {Shimasaku} K.,  {Oguri} M.,  {Ouchi} M.,
  {Tanigawa} S.,  2018, \mn@doi [\apj] {10.3847/1538-4357/aaa6cf}, \href
  {https://ui.adsabs.harvard.edu/abs/2018ApJ...855....4K} {855, 4}

\bibitem[\protect\citeauthoryear{{Kelly} et~al.,}{{Kelly}
  et~al.}{2015}]{Kelly2015}
{Kelly} P.~L.,  et~al., 2015, \mn@doi [Science] {10.1126/science.aaa3350},
  \href {https://ui.adsabs.harvard.edu/abs/2015Sci...347.1123K} {347, 1123}

\bibitem[\protect\citeauthoryear{{Kennedy} \& {Eberhart}}{{Kennedy} \&
  {Eberhart}}{1995}]{Kennedy1995}
{Kennedy} J.,  {Eberhart} R.,  1995, in Proceedings of ICNN'95 - International
  Conference on Neural Networks. pp 1942--1948 vol.4,
  \mn@doi{10.1109/ICNN.1995.488968}

\bibitem[\protect\citeauthoryear{{Kneib} \& {Natarajan}}{{Kneib} \&
  {Natarajan}}{2011}]{Kneib2011}
{Kneib} J.-P.,  {Natarajan} P.,  2011, \mn@doi [\aapr]
  {10.1007/s00159-011-0047-3}, \href
  {http://adsabs.harvard.edu/abs/2011A%26ARv..19...47K} {19, 47}

\bibitem[\protect\citeauthoryear{{Koopmans}}{{Koopmans}}{2005}]{Koopmans2005}
{Koopmans} L.~V.~E.,  2005, \mn@doi [\mnras]
  {10.1111/j.1365-2966.2005.09523.x}, \href
  {http://adsabs.harvard.edu/abs/2005MNRAS.363.1136K} {363, 1136}

\bibitem[\protect\citeauthoryear{{Kormann}, {Schneider}  \&
  {Bartelmann}}{{Kormann} et~al.}{1994}]{Kormann1994}
{Kormann} R.,  {Schneider} P.,   {Bartelmann} M.,  1994, \aap, \href
  {http://adsabs.harvard.edu/abs/1994A%26A...284..285K} {284, 285}

\bibitem[\protect\citeauthoryear{{Liesenborgs}, {de Rijcke}, {Dejonghe}  \&
  {Bekaert}}{{Liesenborgs} et~al.}{2007}]{Liesenborgs2007}
{Liesenborgs} J.,  {de Rijcke} S.,  {Dejonghe} H.,   {Bekaert} P.,  2007,
  \mn@doi [\mnras] {10.1111/j.1365-2966.2007.12236.x}, \href
  {https://ui.adsabs.harvard.edu/abs/2007MNRAS.380.1729L} {380, 1729}

\bibitem[\protect\citeauthoryear{{Limousin} et~al.,}{{Limousin}
  et~al.}{2016}]{Limousin2016}
{Limousin} M.,  et~al., 2016, \mn@doi [Astronomy and Astrophysics]
  {10.1051/0004-6361/201527638}, \href
  {https://ui.adsabs.harvard.edu/abs/2016A&A...588A..99L} {588, A99}

\bibitem[\protect\citeauthoryear{{Lotz} et~al.,}{{Lotz}
  et~al.}{2017}]{Lotz2017}
{Lotz} J.~M.,  et~al., 2017, \mn@doi [\apj] {10.3847/1538-4357/837/1/97}, \href
  {http://adsabs.harvard.edu/abs/2017ApJ...837...97L} {837, 97}

\bibitem[\protect\citeauthoryear{{Marshall} et~al.,}{{Marshall}
  et~al.}{2007}]{Marshall2007}
{Marshall} P.~J.,  et~al., 2007, \mn@doi [\apj] {10.1086/523091}, \href
  {https://ui.adsabs.harvard.edu/abs/2007ApJ...671.1196M} {671, 1196}

\bibitem[\protect\citeauthoryear{{Meneghetti} et~al.,}{{Meneghetti}
  et~al.}{2017}]{Meneghetti2017}
{Meneghetti} M.,  et~al., 2017, \mn@doi [\mnras] {10.1093/mnras/stx2064}, \href
  {http://adsabs.harvard.edu/abs/2017MNRAS.472.3177M} {472, 3177}

\bibitem[\protect\citeauthoryear{{Natarajan} et~al.,}{{Natarajan}
  et~al.}{2017}]{Natarajan2017}
{Natarajan} P.,  et~al., 2017, \mn@doi [\mnras] {10.1093/mnras/stw3385}, \href
  {https://ui.adsabs.harvard.edu/abs/2017MNRAS.468.1962N} {468, 1962}

\bibitem[\protect\citeauthoryear{{Nightingale} \& {Dye}}{{Nightingale} \&
  {Dye}}{2015}]{Nightingale2015}
{Nightingale} J.~W.,  {Dye} S.,  2015, \mn@doi [\mnras]
  {10.1093/mnras/stv1455}, \href
  {https://ui.adsabs.harvard.edu/abs/2015MNRAS.452.2940N} {452, 2940}

\bibitem[\protect\citeauthoryear{{Postman} et~al.,}{{Postman}
  et~al.}{2012}]{Postman2012}
{Postman} M.,  et~al., 2012, \mn@doi [\apjs] {10.1088/0067-0049/199/2/25},
  \href {http://adsabs.harvard.edu/abs/2012ApJS..199...25P} {199, 25}

\bibitem[\protect\citeauthoryear{{Priewe}, {Williams}, {Liesenborgs}, {Coe}  \&
  {Rodney}}{{Priewe} et~al.}{2017}]{Priewe2017}
{Priewe} J.,  {Williams} L.~L.~R.,  {Liesenborgs} J.,  {Coe} D.,   {Rodney}
  S.~A.,  2017, \mn@doi [\mnras] {10.1093/mnras/stw2785}, \href
  {https://ui.adsabs.harvard.edu/abs/2017MNRAS.465.1030P} {465, 1030}

\bibitem[\protect\citeauthoryear{{Rau}, {Vegetti}  \& {White}}{{Rau}
  et~al.}{2014}]{Rau2014}
{Rau} S.,  {Vegetti} S.,   {White} S.~D.~M.,  2014, \mn@doi [\mnras]
  {10.1093/mnras/stu1189}, \href
  {http://adsabs.harvard.edu/abs/2014MNRAS.443..957R} {443, 957}

\bibitem[\protect\citeauthoryear{{Refregier}}{{Refregier}}{2003}]{Refregier2003}
{Refregier} A.,  2003, \mn@doi [\mnras] {10.1046/j.1365-8711.2003.05901.x},
  \href {https://ui.adsabs.harvard.edu/abs/2003MNRAS.338...35R} {338, 35}

\bibitem[\protect\citeauthoryear{{Remolina Gonz{\'a}lez}, {Sharon}  \&
  {Mahler}}{{Remolina Gonz{\'a}lez} et~al.}{2018}]{Remolina2018}
{Remolina Gonz{\'a}lez} J.~D.,  {Sharon} K.,   {Mahler} G.,  2018, \mn@doi
  [\apj] {10.3847/1538-4357/aacf8e}, \href
  {https://ui.adsabs.harvard.edu/abs/2018ApJ...863...60R} {863, 60}

\bibitem[\protect\citeauthoryear{{Richard}, {Pell{\'o}}, {Schaerer}, {Le
  Borgne}  \& {Kneib}}{{Richard} et~al.}{2006}]{Richard2006}
{Richard} J.,  {Pell{\'o}} R.,  {Schaerer} D.,  {Le Borgne} J.~F.,   {Kneib}
  J.~P.,  2006, \mn@doi [\aap] {10.1051/0004-6361:20053724}, \href
  {https://ui.adsabs.harvard.edu/abs/2006A&A...456..861R} {456, 861}

\bibitem[\protect\citeauthoryear{Schmidt et~al.,}{Schmidt
  et~al.}{2014}]{Schmidt2014}
Schmidt K.~B.,  et~al., 2014, \mn@doi [The Astrophysical Journal]
  {10.1088/2041-8205/782/2/l36}, 782, L36

\bibitem[\protect\citeauthoryear{{Sebesta}, {Williams}, {Mohammed}, {Saha}  \&
  {Liesenborgs}}{{Sebesta} et~al.}{2016}]{Williams2016Sebesta}
{Sebesta} K.,  {Williams} L.~L.~R.,  {Mohammed} I.,  {Saha} P.,   {Liesenborgs}
  J.,  2016, \mn@doi [\mnras] {10.1093/mnras/stw1433}, \href
  {https://ui.adsabs.harvard.edu/abs/2016MNRAS.461.2126S} {461, 2126}

\bibitem[\protect\citeauthoryear{{Seitz} \& {Schneider}}{{Seitz} \&
  {Schneider}}{1997}]{Seitz1997}
{Seitz} C.,  {Schneider} P.,  1997, \aap, \href
  {https://ui.adsabs.harvard.edu/abs/1997A&A...318..687S} {318, 687}

\bibitem[\protect\citeauthoryear{{Sersic}}{{Sersic}}{1968}]{Sersic1968}
{Sersic} J.~L.,  1968, {Atlas de Galaxias Australes}

\bibitem[\protect\citeauthoryear{Shajib et~al.,}{Shajib
  et~al.}{2018}]{Shajib2019}
Shajib A.~J.,  et~al., 2018, \mn@doi [Monthly Notices of the Royal Astronomical
  Society] {10.1093/mnras/sty3397}, 483, 5649

\bibitem[\protect\citeauthoryear{{Shajib} et~al.,}{{Shajib}
  et~al.}{2019}]{Shajib:2019H0}
{Shajib} A.~J.,  et~al., 2019, arXiv e-prints, \href
  {https://ui.adsabs.harvard.edu/abs/2019arXiv191006306S} {p. arXiv:1910.06306}

\bibitem[\protect\citeauthoryear{{Sharon} \& {Johnson}}{{Sharon} \&
  {Johnson}}{2015}]{Sharon2015}
{Sharon} K.,  {Johnson} T.~L.,  2015, \mn@doi [\apjl]
  {10.1088/2041-8205/800/2/L26}, \href
  {http://adsabs.harvard.edu/abs/2015ApJ...800L..26S} {800, L26}

\bibitem[\protect\citeauthoryear{{Sharon}, {Gladders}, {Rigby}, {Wuyts},
  {Koester}, {Bayliss}  \& {Barrientos}}{{Sharon} et~al.}{2012}]{Sharon2012}
{Sharon} K.,  {Gladders} M.~D.,  {Rigby} J.~R.,  {Wuyts} E.,  {Koester} B.~P.,
  {Bayliss} M.~B.,   {Barrientos} L.~F.,  2012, \mn@doi [\apj]
  {10.1088/0004-637X/746/2/161}, \href
  {http://adsabs.harvard.edu/abs/2012ApJ...746..161S} {746, 161}

\bibitem[\protect\citeauthoryear{{Stark}, {Swinbank}, {Ellis}, {Dye}, {Smail}
  \& {Richard}}{{Stark} et~al.}{2008}]{Stark2008}
{Stark} D.~P.,  {Swinbank} A.~M.,  {Ellis} R.~S.,  {Dye} S.,  {Smail} I.~R.,
  {Richard} J.,  2008, \mn@doi [\nat] {10.1038/nature07294}, \href
  {https://ui.adsabs.harvard.edu/abs/2008Natur.455..775S} {455, 775}

\bibitem[\protect\citeauthoryear{{Suyu}, {Marshall}, {Blandford}, {Fassnacht},
  {Koopmans}, {McKean}  \& {Treu}}{{Suyu} et~al.}{2009}]{Suyu2009}
{Suyu} S.~H.,  {Marshall} P.~J.,  {Blandford} R.~D.,  {Fassnacht} C.~D.,
  {Koopmans} L.~V.~E.,  {McKean} J.~P.,   {Treu} T.,  2009, \mn@doi [\apj]
  {10.1088/0004-637X/691/1/277}, \href
  {http://adsabs.harvard.edu/abs/2009ApJ...691..277S} {691, 277}

\bibitem[\protect\citeauthoryear{{Tagore} \& {Keeton}}{{Tagore} \&
  {Keeton}}{2014}]{Tagore2014}
{Tagore} A.~S.,  {Keeton} C.~R.,  2014, \mn@doi [\mnras]
  {10.1093/mnras/stu1671}, \href
  {http://adsabs.harvard.edu/abs/2014MNRAS.445..694T} {445, 694}

\bibitem[\protect\citeauthoryear{{Treu}}{{Treu}}{2010}]{Treu2010}
{Treu} T.,  2010, \mn@doi [\araa] {10.1146/annurev-astro-081309-130924}, \href
  {http://adsabs.harvard.edu/abs/2010ARA%26A..48...87T} {48, 87}

\bibitem[\protect\citeauthoryear{{Treu} \& {Koopmans}}{{Treu} \&
  {Koopmans}}{2004}]{Treu2004}
{Treu} T.,  {Koopmans} L.~V.~E.,  2004, \mn@doi [\apj] {10.1086/422245}, \href
  {http://adsabs.harvard.edu/abs/2004ApJ...611..739T} {611, 739}

\bibitem[\protect\citeauthoryear{Treu et~al.,}{Treu et~al.}{2015}]{Treu2015}
Treu T.,  et~al., 2015, \mn@doi [The Astrophysical Journal]
  {10.1088/0004-637x/812/2/114}, 812, 114

\bibitem[\protect\citeauthoryear{{Vegetti} \& {Koopmans}}{{Vegetti} \&
  {Koopmans}}{2009}]{Vegetti2009}
{Vegetti} S.,  {Koopmans} L.~V.~E.,  2009, \mn@doi [\mnras]
  {10.1111/j.1365-2966.2008.14005.x}, \href
  {http://adsabs.harvard.edu/abs/2009MNRAS.392..945V} {392, 945}

\bibitem[\protect\citeauthoryear{{Wagner}}{{Wagner}}{2019}]{Wagner2019}
{Wagner} J.,  2019, \mn@doi [Universe] {10.3390/universe5070177}, \href
  {https://ui.adsabs.harvard.edu/abs/2019Univ....5..177W} {5, 177}

\bibitem[\protect\citeauthoryear{{Warren} \& {Dye}}{{Warren} \&
  {Dye}}{2003}]{Warren2003}
{Warren} S.~J.,  {Dye} S.,  2003, \mn@doi [\apj] {10.1086/375132}, \href
  {https://ui.adsabs.harvard.edu/abs/2003ApJ...590..673W} {590, 673}

\bibitem[\protect\citeauthoryear{Zitrin et~al.,}{Zitrin
  et~al.}{2015}]{Zitrin2015}
Zitrin A.,  et~al., 2015, \mn@doi [The Astrophysical Journal]
  {10.1088/0004-637x/801/1/44}, 801, 44

\bibitem[\protect\citeauthoryear{{de La Vieuville} et~al.,}{{de La Vieuville}
  et~al.}{2019}]{deLaVieuville2019}
{de La Vieuville} G.,  et~al., 2019, \mn@doi [\aap]
  {10.1051/0004-6361/201834471}, \href
  {https://ui.adsabs.harvard.edu/abs/2019A&A...628A...3D} {628, A3}

\makeatother
\end{thebibliography}





\bsp	
\label{lastpage}
\end{document}